\documentclass[useAMS,usenatbib]{mn2e}
\pdfoutput=1

\usepackage{graphicx}
\usepackage{subfigure}
 \usepackage{graphicx,amsmath,amssymb}
\usepackage{color} 
\usepackage{hyperref}
\usepackage{color}
\usepackage{cleveref}
\def\lsim{\lower.5ex\hbox{$\; \buildrel < \over \sim \;$}}
\def\gsim{\lower.5ex\hbox{$\; \buildrel > \over \sim \;$}}
\def\simeq{\lower.3ex\hbox{$\; \buildrel \sim \over - \;$}}
\def\ch{\lower-0.55ex\hbox{--}\kern-0.55em{\lower0.15ex\hbox{$h$}}}
\def\lh{\lower-0.55ex\hbox{--}\kern-0.55em{\lower0.15ex\hbox{$\lambda$}}}

\def\etal{{\em et al.} }

\def\etal{{\em et al.}}

\def\lsim{\lower.5ex\hbox{$\; \buildrel < \over \sim \;$}}
\def\gsim{\lower.5ex\hbox{$\; \buildrel > \over \sim \;$}}
\def \simeq{\lower.3ex\hbox{$\; \buildrel \sim \over - \;$}}

\def\Nature{{Nature}}%
\def\araa{{ARA\&A}}%
\def\apj{{ApJ}}%
\def\apjl{{ApJL}}%
\def\aap{{A\&A}}%
\def\mnras{{MNRAS}}%
\def\na{{New Astron.}}%
\def\nar{{New A Rev.}}%
\def\pasj{{PASJ}}%
\def\NCRS{{Nuovo Cimento Rivista Serie}}%
\def\asinc{{BASI}}%

\title[Estimation of mass outflow rates from dissipative accretion disc
around rotating black holes]
{Estimation of mass outflow rates from dissipative accretion disc
around rotating black holes}
\author[Ramiz Aktar, Santabrata Das, Anuj Nandi, H. Sreehari]
{Ramiz Aktar$^{1}$\thanks{E-mail:
ramiz@iitg.ernet.in (RA); sbdas@iitg.ernet.in (SD); anuj@isac.gov.in (AN); sreehari@physics.iisc.ernet.in (HS)}, 
Santabrata Das$^{1}$, Anuj Nandi$^{2}$, H. Sreehari$^{2,3}$
\footnotemark[1] \\ 
$^{1}$Indian Institute of Technology Guwahati, Guwahati, 781039, India\\
$^{2}$Space Astronomy Group, ISITE Campus, ISRO Satellite Centre, Outer
Ring Road, Marathahalli, Bangalore, 560037, India\\
$^{3}$Indian Institute of Science, Bangalore, 560012, India}

\begin{document}

\date{Accepted . Received ; in original form }

\pagerange{\pageref{firstpage}--\pageref{lastpage}} \pubyear{}

\maketitle

\label{firstpage}

\begin{abstract}

We study the properties of the dissipative accretion flow around rotating
black holes in presence of mass loss. We obtain the complete set of global
inflow-outflow solutions in the steady state by solving the underlying
conservation equations self-consistently. We observe that global inflow-outflow
solutions are not the isolated solution, instead such solutions are possible
for wide range of inflow parameters. Accordingly, we identify the boundary
of the parameter space for outflows, spanned by the angular momentum
($\lambda_{\rm in}$) and the energy (${\cal E}_{\rm in}$) at the inner
sonic point ($x_{\rm in}$), as function of the dissipation parameters
and find that parameter space gradually shrinks with the increase of
dissipation rates. Further, we examine the properties of the outflow
rate $R_{\dot m}$ (defined as the ratio of outflow to inflow mass flux)
and ascertain that dissipative processes play the decisive role in determining
the outflow rates. We calculate the limits on the maximum outflow rate
($R_{\dot{m}}^{\rm max}$) in terms of viscosity parameter ($\alpha$) as
well as black hole spin ($a_k$) and obtain the limiting range as
$3\% \le R_{\dot{m}}^{\rm max} \le 19\%$. Moreover, we calculate the viable
range of $\alpha$ that admits the coupled inflow-outflow solutions and find that
$\alpha \lesssim 0.25$ for $R_{\dot m} \ne 0$. Finally, we discuss the
observational implication of our formalism to infer the spin of the black holes. 
Towards this, considering the highest observed QPO frequency of black hole source
GRO J1655-40 ($\sim 450$ Hz), we constrain the spin value of the source as
$a_k \ge 0.57$.

\end{abstract}


\section{Introduction}

In accretion system around black holes, jets and outflows play an essential role
and their existence is realized both in observations and simulations
\citep{Mirabel-etal92,Mirabel-Rodriguez94, Hjellming-Rupen95,Ferrari98,
Mirabel-Rodriguez98,Junor-etal99,Cheung02,Mirabel03,Fender-etal09,
Miller-etal12b,Mezcua-etal13,Das-etal14,Okuda-Das15,Mezcua-etal15,
Fender-Munoz-Darias16}. Since black holes do not have any hard
boundary, the outflows and jets must be originated from the
accretion disc and observations also indeed confirm the disc-jet
connection \citep{Feroci-99,Vadawale-etal01,
Nandi-etal01,Gallo-etal03, Miller-etal12a, MillerJones-etal12, 
Corbel-etal13, Radhika-Nandi14, Radhika-etal16a}. 
Earlier theoretical works of \citet{Penrose69}
and \citet{Blandford-Znajek77} suggested that the powerful jets are
originated due to the rotation of the black hole. However, conflicting claims
were made from the observational front.
\citet{Steiner-etal13,McClintock-etal14}
reported the significant evidences of  positive correlation between the jet
power and the spin of the black holes. On the other hand, \citet{Russell-etal13,
Fender-Gallo14} did not find any such correlation in their study. Very recently,
\citet{Aktar-etal15} calculated the mass loss from inviscid advective 
disc considering accretion-ejection
model and found a feeble correlation between the maximum outflow rates and
the spin of the black holes. This result possibly indicates that the existence
of jet-spin correlation in the accretion-ejection system seems to be elusive.

\clearpage
Extensive efforts were made to infer the exact mechanism of
jet generation and the powering of jets 
\citep{Nandi-etal01,Falcke-etal04,Fender-etal04,
Fender-etal09, Steiner-etal13,Russell-etal13,McClintock-etal14,Fender-Gallo14}.
Several attempts were pursued to investigate the accretion-ejection
coupling mechanism considering the inflowing matter to be sub-Keplerian in
nature. In these works, both the accretion and ejection processes are
overall governed by the conservation laws of gas dynamics, namely
the conservation of mass, momentum and energy. In reality, during the
course of the accretion process, subsonic rotating flow starts it journey
from the outer edge of the disc with negligible radial velocity. Due
to the influence of strong gravity, subsonic matter gradually gains
its radial velocity while moving towards the black hole and becomes 
super sonic after crossing the sonic point. Depending on the angular
momentum, accretion flow may contain multiple sonic points and in this
case, accreting matter after passing through the outer sonic point
experiences centrifugal barrier while moving further towards the
horizon. This causes the piling of matter which eventually may trigger the
shock transition. The feasibility of shock solutions and its implications
have been studied extensively around non-rotating black holes \citep{Fukue87,
Chakrabarti89,Lu-etal99,Becker-Kazanas01,Das-etal01a,Fukumura-Tsuruta04,
Chakrabarti-Das04,Das07,Das-etal09,Sarkar-Das16} as well as rotating black holes
\citep{Chakrabarti96b,Mondal-Chakrabarti06,Das-Chakrbarti08,Das-etal10,Aktar-etal15}.

The advantage of the shock induced accretion solution is that due to
shock compression, the post shock region of the accretion disc around
black holes becomes hot and dense and eventually, innermost region of
the disc does not remain confined around the disc equatorial plane,
instead it becomes puffed up. Apparently, the post shock flow around the
black holes behaves like a Comptonizing cloud (equivalently Post Shock Corona,
hereafter PSC). During accretion, inflowing matter experiences excess
thermal gradient force at PSC which drives a part of the infalling
matter in the vertical direction to produce bipolar outflows. This
underlying mechanism possibly establishes the coupling of outflow
generation with the accretion dynamics. Extensive numerical simulations
in hydrodynamics (HD) as well as magnetohydrodynamics (MHD) environment
confirm the disk-jet connection where the generation of mass outflow
takes place from the inner part of the disc \citep{Molteni-etal94,
Molteni-etal96a, Machida-etal00, Koide-etal02,McKinney-Gammie04,
De-Villiers-etal05,Giri-etal10,Okuda14, Das-etal14,Okuda-Das15} 

Considering the above appealing mechanism of outflow formation,
\citet{Chakrabarti99} and \citet{Das-etal01a} estimated the rate of
mass loss from the disc self-consistently for adiabatic flow. Subsequently,
the studies of outflow rates from accretion flows are demonstrated with
variety and complexity of disc-jet structures \citep{Das-etal01a,
Chattopadhyay-Das07,Das-Chattopadhyay08,Kumar-Chattopadhyay13,Aktar-etal15,
Chattopadhyay-Kumar16}.
\citet{Das-Chattopadhyay08} performed a detail study of mass loss for
dissipative steady accretion flows around non-rotating black holes
including the effects of viscosity and Synchrotron cooling processes. Recently,
\citet{Chattopadhyay-Kumar16} further examined accretion-ejection
solutions around Schwarzschild black hole in full general relativity.
Meanwhile, \citet{Aktar-etal15} investigated the outflow properties around
the rotating black holes for inviscid flow. In these studies, the effects of
dissipative processes and the black hole rotation on the accretion-ejection
solutions are carried out separately. It is therefore pertinent to investigate
how the properties of the accretion as well as the outflow solutions depend
on the black hole spin and the dissipation parameters which is apparently the
main purpose of this work. 

Accordingly, in this paper, we undertake a steady, axisymmetric, sub-Keplerian,
viscous advective accretion flow around a rotating black hole. Indeed, as
\citet{Chakrabarti96c} pointed out that the 
sub-Keplerian accretion around central black hole seems to be 
viable, when matter is accreted from the winds of the surrounding stars having
very little angular momentum. Moreover, following the work of 
\citet{Narayan-Yi94}, here we introduce the effect of cooling in a parametric way.
To avoid the complexities of full general relativistic
calculations, we adopt the pseudo-Kerr potential \citep{Mondal-Chakrabarti06}
that satisfactorily describes the space-time geometry around the rotating
black holes having spin parameter $a_k \lesssim 0.8$.
In this problem, the bipolar jets and outflows are 
assumed to be launched from the PSC and the jets are assumed to be confined
within the physically motivated jet geometry which is constructed based on
the treatment proposed by \citet{Molteni-etal96a}. We calculate the shock
induced global transonic accretion solutions in presence and absence
of thermally driven outflows and identify the parameter spaces spanned by
the local energy (${\cal E}_{\rm in}$) and angular momentum ($\lambda_{\rm in}$)
at the inner sonic point ($x_{\rm in}$) of the inflowing mater in terms
of the black hole spin. In addition, we also performed the classification
of the parameter spaces as function of the viscosity and cooling parameters.
We find that the available parameter space for shock is always reduced when
outflow is present.
It is observed that the accretion-ejection solutions exist for a wide range
of the inflow parameters. Furthermore, we compute the maximum mass outflow
rate ($R_{\dot m}^{\rm max}$) in terms of viscosity parameter ($\alpha$) by varying
the inflow parameters and observed that $R_{\dot m}^{\rm max}$ is predominantly
obtained for higher ${\cal E}_{\rm in}$ and lower $\lambda_{\rm in}$ values
irrespective to the spin of the black holes. Next, we calculate the maximum
values of the viscosity parameter ($\alpha^{\rm max}$) as function of black
hole spin ($a_k$) both by excluding outflow ($\alpha_{\rm no}^{\rm max}$)
and including outflows ($\alpha_{\rm o}^{\rm max}$). Since the shock
parameter space is shrunk when outflow is included, for a given $a_k$, 
we obtain $\alpha_{\rm o}^{\rm max} < \alpha_{\rm no}^{\rm max}$ all throughout.
Moreover, in our present analysis, we find that the estimated limits of
$\alpha_{\rm no}^{\rm max} \sim 0.35$ and $\alpha_{\rm o}^{\rm max} \sim 0.25$
for non-rotating black hole ($a_k= 0$) are very much consistent with the
results reported by \citet{Chakrabarti-Das04,Kumar-Chattopadhyay13}.
Following our present inflow-outflow formalism, we attempt to find out the
plausible origin of the high frequency QPOs and also explore the possibility
of constraining the spin of the black hole sources. As a representative case, 
we  analyze the {\it RXTE$^1$} \footnotetext[1]{https://heasarc.gsfc.nasa.gov/docs/xte/xtegof.html}
satellite archival data of a particular
observation of the black hole source GRO J1655-40 as it exhibits the highest
observed QPO frequency ($\sim 450~{\rm Hz}$) among all the known black hole sources.
Based on the present approach, we argue that the spin of GRO J1655-40 seems to be
$a_k\ge0.57$.

The remainder of the paper is organized as follows. In \S 2, we
describe the governing equations for accretion and outflows. In the next
section (\S 3), we discuss the solution methodology to calculate
the inflow-outflow
solutions in terms of the inflow parameters and estimate the mass outflow rates.
In \S 4, we present the results in detail and discuss the possible observational
implications of our formalism to constrain the spin of the black hole
sources. Finally, we review our main conclusions in \S 5.

\section{Model Equations and Assumptions}

In this article, we consider a disc-jet system around a rotating black hole where
we assume that the accretion disc lies along the equatorial plane and the jet geometry
is considered in the off-equatorial plane about the black hole rotation axis
\citep[see Fig. 1]{Molteni-etal96a,Aktar-etal15}. To represent the flow variables,
we use the geometric unit system as $G$ = $M_{\rm BH}$ = $c$ = 1, where, $G$, $M_{\rm BH}$ and 
$c$ are the Gravitational constant, mass of the black hole and speed of light,
respectively.

\subsection{Governing Equations for Accretion}

We consider a steady, geometrically thin, axisymmetric, viscous accretion
flow around a rotating black hole. To avoid the complexity of full
general relativistic calculation, we adopt the pseudo-Kerr potential approach 
to mimic the space-time geometry around a rotating black hole. The equations of 
motion for accreting matter are given by,

(i) The radial momentum equation:
$$
v\frac{dv}{dx} + \frac{1}{\rho}\frac{dP}{dx} + \frac{d\Phi_{eff}}{dx} =0 ,
\eqno(1)
$$ 
where, $v$, $P$ and $\rho$ represent the radial velocity, gas pressure and
 density of the accretion flow respectively. Here, $\Phi_{eff}$ denotes the pseudo-Kerr 
effective potential proposed by \citet{Chakrabarti-Mondal06}. The expression of  $\Phi_{eff}$ is given by,
$$
\Phi_{eff}=-\frac{\mathcal{B}+\sqrt{\mathcal{B}^2-4\mathcal{A}\mathcal{C}}}{2\mathcal{A}},
\eqno(2)
$$
where,
\begin{align*}
\mathcal{A} &=\frac{\epsilon^2 \lambda^2}{2x^2},\\
\mathcal{B} &=-1 + \frac{\epsilon^2 \omega \lambda r^2}{x^2} 
+\frac{2a_k\lambda}{r^2 x},\\
\mathcal{C} &=1-\frac{1}{r-x_0}+\frac{2a_k\omega}{x}
+\frac{\epsilon^2 \omega^2 r^4}{2x^2}.
\end{align*}
Here, $x$ and $r$ represent the cylindrical and spherical radial distance
considering the black hole is located at the origin of the coordinate system
and $\lambda$ is the specific angular momentum of the flow.
Here, $x_0=0.04+0.97a_k+0.085a_k^2$, $\omega=2a_k/(x^3+a^2_k 
x+2a^2_k)$
and $\epsilon^2=(x^2-2x+a^2_k)/(x^2+a_k^2+2a^2_k/x)$, $\epsilon$ is the
redshift factor and $a_k$ represents the black hole rotation parameter
defined as the specific spin angular momentum of the black hole. 
According to \citet{Chakrabarti-Mondal06}, the above potential mimic the Kerr 
geometry quite satisfactorily for $a_k \leq 0.8$.

(ii) The mass conservation equation:
$$
\dot{M}=4 \pi \rho v x h ,
\eqno(3)
$$
where, ${\dot M}$ denotes the mass accretion rate which is constant everywhere
except the region of mass loss and $4\pi$ is the geometric constant. Considering the
hydrostatic equilibrium in the vertical direction, the half-thickness of the disc $h(x)$
is obtained as,
$$
h(x)=a\sqrt{\frac{x}{\gamma \Phi_{r}^{\prime}}},
\eqno(4)
$$
where, $a$ is the adiabatic sound speed defined as $a=\sqrt{\frac{\gamma P}
{\rho}}$ and $\gamma$ is the adiabatic index. Here, $\Phi_{r}^{\prime}=\left( 
\frac{\partial \Phi_{eff}}{\partial r}\right)_{z<<x}$, and $z$ is
the vertical height in the cylindrical coordinate system where $r = \sqrt{x^2 + z^2}$
\citep{Das-etal10}.

(iii) The angular momentum distribution equation:
$$
v\frac{d\lambda}{dx} + \frac{1}{\Sigma x}\frac{d}{dx}(x^2 W_{x\phi}) =0.
\eqno(5)
$$
Here, we assume that the viscous stress is dominated by the
$x\phi$ component and is denoted by $W_{x\phi}$. In this work, for the sake of
completeness, we consider the viscous stress which is given by \citet{Chakrabarti96a},

$$
W_{x\phi}^{(1)} = -\alpha (W + \Sigma v^2),
\eqno(6)
$$
where,  $W = 2I_{n+1}Ph(x)$ and $\Sigma = 2 I_n \rho h(x)$ are the vertically
integrated pressure and density, respectively \citep{moto84} and $\alpha$
refers to the viscosity parameter. When radial velocity of the inflowing matter
is insignificant as in the case of Keplerian disc, equation (6) reduces to the
seminal viscosity prescription of \citet{SS73}.
It is to be noted that an alternative expression of the viscous shear stress is also considered
while modeling the accretion flow around black hole
\citep[and reference therein]{Chakrabarti96a} which is given by,
$$
W_{x\phi}^{(2)} = \eta x \frac{d\Omega}{dx},
\eqno(7)
$$
where, $\Omega(x)$ is the angular velocity of the accreting matter.

\noindent And finally,

(iv) The entropy generation equation:
$$
\Sigma v T \frac{ds}{dx} = Q^{+} - Q^{-} = fQ^{+}; f = 1 - \frac{Q^{-}}{Q^{+}},
\eqno(8)
$$
where, $T$, and $s$ are the temperature and entropy density of the 
accretion flow, respectively. 
Here, $Q^{+}$ and $Q^{-}$ denote the heat gain and lost by the flow. 
Here, we introduce a parametric cooling factor $f$ in our calculation
following the work of \citet{Narayan-Yi94}, where the value of $f$ lies in the
range $0\leq f \leq 1$. When $f = 1$, accretion flow behaves like advection
dominated whereas for $f = 0$, flow becomes cooling dominated.

After some simple algebra, equation (8) yields as,
$$
\frac{v}{\gamma -1}\left[\frac{1}{\rho}\frac{dP}{dx} - \frac{\gamma P}
{\rho^2}\frac{d\rho}{dx}\right]=-\frac{fQ^{+}}{\rho h}=-H.
\eqno(9)
$$
where, we define $H=\frac{fQ^{+}}{\rho h}$.
Due to viscous shear, accreting matter is heated up and we
compute the heating of the flow as
$Q^{+}=W_{x\phi}^{2}/\eta$ where $\eta$ represents the dynamical viscosity
coefficient. To calculate $Q^{+}$, when $W_{x\phi}^{(1)}$ is used, the contribution
of viscous shear is being sacrificed. On the other hand, when $W_{x\phi}^{(2)}$ is used,
second order derivative of $\Omega$ appeared in the sonic point analysis and equations
become difficult to solve. To avoid such limitations, however, retaining the memory of
$d\Omega/dx$ intact, we adopt the mixed shear stress prescription 
\citep{Chakrabarti96a} where the combination of both shear stresses are considered.
In the present analysis, therefore the heating term $Q^{+}$ yields as,
$$
Q^{+} = \frac{W_{x\phi}^{(1)} W_{x\phi}^{(2)}}{\eta}=- \alpha(W + \Sigma v^2)\left(x\frac{d\Omega}
{dx}\right).
\eqno(10)
$$
Using equation (10), we obtain the simplified expression of $H$ as,
$$
H = \frac{fQ^{+}}{\rho h}= f A(ga^2 + \gamma v^2)\left(x\frac{d\Omega}{dx}\right),
\eqno(11)
$$
where, $A=-2\alpha I_{n}/\gamma$ and $g=I_{n+1}/I_{n}$. 
Here, $n = (\gamma -1)^{-1}$ is the polytropic index. $I_{n}$ and 
$I_{n+1}$ are the constant factors arising while carrying out the vertical integration of
density and pressure at a given radial coordinate $x$ (see \citet{moto84}).

\subsection{Sonic point conditions}

In the accretion process around black hole, radial velocity of the inflowing matter
at the outer edge remains negligibly small although matter enters into the black hole
with radial velocity equal to the speed of light. Therefore, the velocity of the inflowing
matter is likely to match with the sound speed at some point where flow changes its 
sonic character from subsonic to supersonic state. Such a point is called
as sonic point. Depending on the initial parameters, flow may contain more than
one sonic points. Flow containing multiple sonic points is of our interest as
it may possess stationary shock when the shock conditions are satisfied. In order
to investigate the properties of the shock wave and its implication for black hole
accretion, study of sonic point properties is therefore essential.  Accordingly, we carry
out sonic point analysis following \citet{Chakrabarti89} and obtain the first order
linear differential equation using equations (1), (3), (5) and (9)  as,
$$
\frac{dv}{dx} = \frac{N}{D},
\eqno(12)
$$
where,
\begin{align*}
N =& -\frac{f A\alpha  (ga^2 + \gamma v^2)^2}{\gamma  vx} - \frac{3 a^2 v}{(\gamma -1)x} +\frac{ a^2 v}{(\gamma 
-1)}\left(\frac{d\ln\Phi_{r}^{\prime}}{dx}\right) \\ &+ \left[\frac{2fA\alpha g  (ga^2 + \gamma v^2)}{v} + 
\frac{(\gamma +1)v}{(\gamma -1)}\right]\left(\frac{d\Phi_{eff}}
{dx}\right)  
\nonumber \\& - \frac{3f A \alpha g a^2 
(ga^2 + \gamma v^2)}{\gamma vx} \nonumber + \frac{f A \alpha g a^2(ga^2 + \gamma v^2)}
{\gamma v}\left(\frac{d\ln\Phi_{r}^{\prime}}{dx}\right) 
\\& + \frac{2f A\lambda (ga^2 + \gamma v^2)}
{x^2}, \tag{13}
\end{align*}
and 
$$
D = \frac{2 a^2}{(\gamma -1)} - \frac{(\gamma + 1)v^2}{(\gamma - 1)} - 
fA \alpha (ga^2 + \gamma v^2) \left[(2 g - 1) - \frac{g a^2}{\gamma v^2}\right].
\eqno(14)
$$

We calculate the differential form of angular momentum using the 
expression of equation (5) as,
$$
\frac{d\lambda}{dx} = \frac{\alpha}{\gamma v}(ga^2 + \gamma v^2) + 
\frac{2\alpha x ga}{\gamma v}\left(\frac{da}{dx}\right) + \alpha x \left(1 
- \frac{ga^2}{\gamma v^2}\right)\left(\frac{dv}{dx}\right),
\eqno(15)
$$
and also we calculate the gradient of sound speed using equations (1-4) as,
$$
\frac{da}{dx} = \left(\frac{a}{v}-\frac{\gamma v}{a}\right)\frac{dv}{dx} + 
\frac{3a}{2x} - \frac{a}{2}\left(\frac{d\ln\Phi_{r}^{\prime}}{dx}\right) - 
\frac{\gamma}{a}\left(\frac{d\Phi_{eff}}{dx}\right).
\eqno(16)
$$

The inner boundary condition of black hole demands that the flow
should be smooth everywhere between the outer edge and the 
horizon.  Therefore, the radial velocity gradient must be finite
everywhere. To maintain this, at the sonic point, both 
numerator $N$ and denominator $D$ of Eq. (12) must vanish 
simultaneously \citep{Chakrabarti89}. Setting $D =0$, we obtain the
Mach number ($M =v/a$)  expression at the sonic point as,

$$
M^2_c = \frac{-m_b - \sqrt{m_b^2 - 4 m_a m_c}}{2 m_a},
\eqno(17)
$$
where, 
\begin{align*}
m_a =&-fA\alpha \gamma^2 (\gamma - 1)(2g - 1) - \gamma (\gamma +1), \tag{17a}\\ \tag{17b}
 m_b =& 2\gamma - 2fA \alpha g \gamma (\gamma -1)(g - 1),\\ \tag{17c}
 m_c =&fA\alpha g^2(\gamma -1).
\end{align*}

The sound speed at the sonic point is found out by setting numerator $N=0$ and is given by,

$$
a(x_c) = \frac{-a_2-\sqrt{a^2_2 - 4a_1 a_3}}{2a_1},
\eqno(18)
$$
where,
\begin{align*}
a_1 =& -\frac{fA\alpha (g + \gamma M_c^2)^2}{\gamma x} - \frac{3 M_c^2}{(\gamma -1)x} 
+ \frac{M_c^2}{(\gamma -1)}\left(\frac{d\ln\Phi_{r}^{\prime}}{dx}\right)
\\ & - \frac{3fA \alpha g (g + \gamma M_c^2)}{\gamma x}   + \frac{fA \alpha g(g + \gamma M_c^2)}
{\gamma}\left(\frac{d\ln\Phi_{r}^{\prime}}{dx}\right), \tag{18a} \\
a_2 =&  \frac{2fA\lambda M_c (g + \gamma M_c^2)}
{x^2}, \tag{18b} \\
a_3 =& \left[2fA\alpha g (g + \gamma M_c^2) + \frac{(\gamma + 1)M_c^2}{(\gamma - 
1)}\right]\left(\frac{d\Phi_{eff}}{dx}\right). \tag{18c}                
\end{align*}
Here, $M_c$ represents the Mach number at the sonic point.

\begin{figure}
\begin{center}
\includegraphics[width=0.45\textwidth]{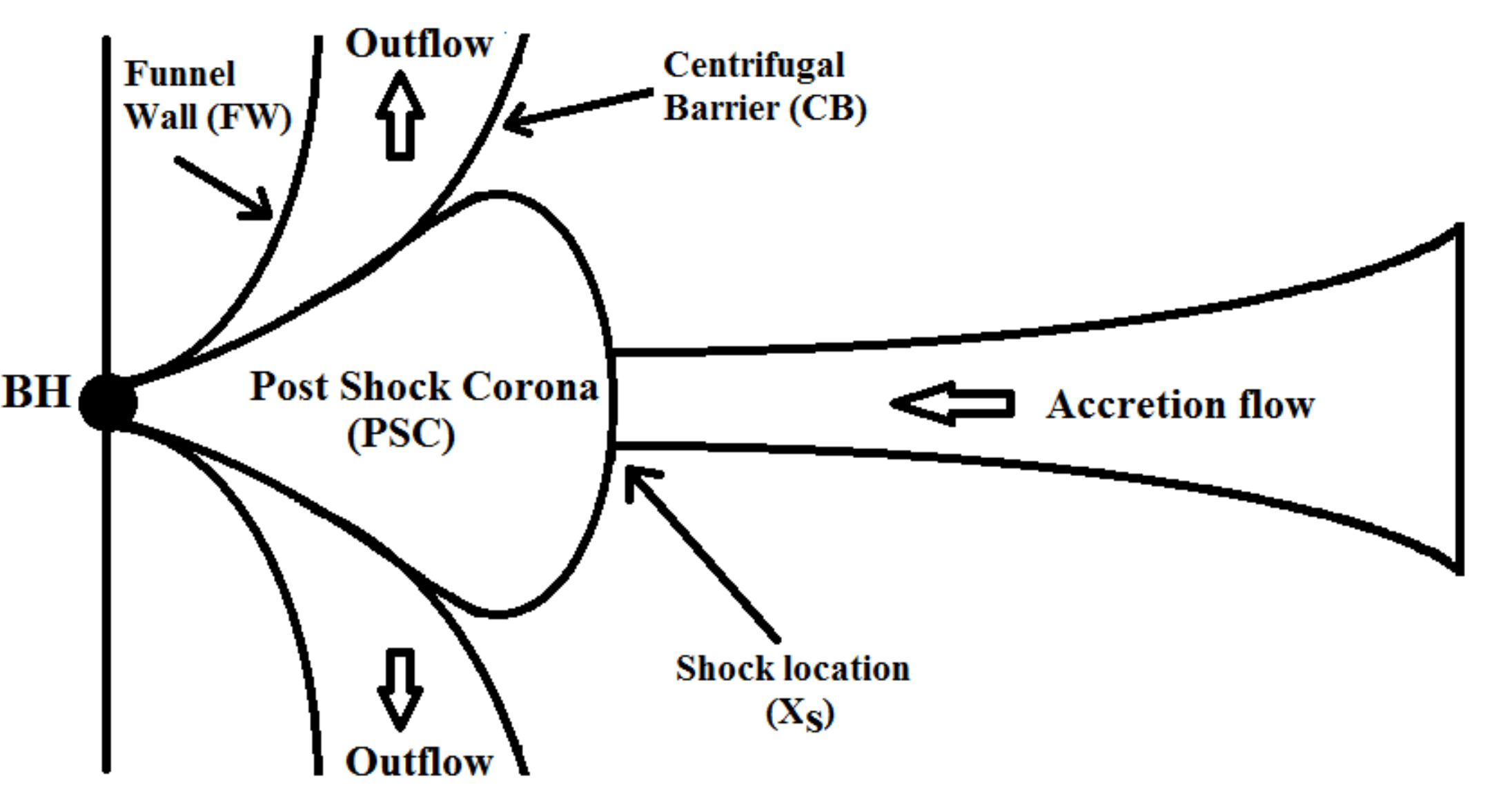}
\end{center}
\caption{Schematic diagram of disc-jet geometry is shown. 
Accretion flow, post-shock corona (PSC) and outflow are depicted in the
figure. Also, we show the shock location ($x_s$), centrifugal barrier
(CB) and funnel wall (FW). Here, the central black hole is indicated
as BH. See text for details.}
\end{figure}

\subsection{Equations for Outflow}

In this work, we consider that the accretion flow geometry is resided 
around the equatorial 
plane and the jet or outflow geometry is described in the off-equatorial 
plane about the axis of rotation of the black hole \citep{Molteni-etal96a,Chattopadhyay-Das07}.
As the jets are tenuous in nature, the differential rotation of the outflowing 
matter is expected to be negligibly small and thus, we ignore the viscosity while
describing outflow. Further, the outflowing matter is considered to obey the 
polytropic equation of state as $P_{j}=K_{j} \rho_{j}^{\gamma}$, where, the 
suffix `$j$' denotes the outflow variables and 
$K_{j}$ denotes the measure of specific entropy of the jet.
Subsequently, the equations of motion of the outflowing matter are given by,

(i) The energy conservation equation of outflow:
$$
\mathcal{E}_{j}=\frac{1}{2}v_j^2+\frac{a_j^2}{\gamma -1 }+\Phi_{eff},
\eqno(19)
$$
where, $\mathcal{E}_{j}$ represents the specific 
energy of the outflow, $v_j$ is the outflow velocity and $a_j$ is the sound 
speed of the outflow, respectively.

(ii) the mass conservation equation of outflow:
$$
{\dot{M}_{out}}=\rho_{j}v_{j}\mathcal{A}_j,
\eqno(20)
$$
where, $\dot{M}_{out}$ represents the outflowing mass rate and $\mathcal{A}_j$
refers to the area function of the jet. We estimate $\mathcal{A}_j$ by using the 
radius of two boundary surfaces, namely the centrifugal barrier (CB) and the 
funnel wall (FW) \citep{Molteni-etal96a} (see Fig. 1). 
The centrifugal barrier (CB) is 
obtained by defining the pressure maxima surface as $(d\Phi^{eff}/dx)_{r_{CB}}=0$ 
and the funnel wall (FW) stands for the pressure minimum surface which is defined 
by the null effective potential as $\Phi^{eff}\vert_{r_{FW}}=0$
 \citep{Molteni-etal96a}.
In general, the jet streamlines within the jet geometry are not exactly parallel to the jet
area vector and therefore, in order to calculate the jet area function, we introduce a geometric
factor corresponding to the projection of the jet streamlines  on the jet cross-section.
Accordingly, the jet area function is obtained as \citep{Kumar-Chattopadhyay13},
$$
\mathcal{A}_j = \frac{2 \pi (x_{CB}^2 - x_{FW}^2)}{\sqrt{1 + 
(dx_j/dy_j)^2}},
\eqno(21)
$$
where, $\sqrt{1 + (dx_j/dy_j)^2}$ is projection factor of the jet area.
Here, $x_{CB}$ and $x_{FW}$ refer to the radius of the
centrifugal barrier and funnel wall in the cylindrical coordinate system, respectively.
The corresponding spherical radius of the jet is given by $r_j = \sqrt{x_j^2 + y_j^2}$.
With this consideration, we carry out the sonic point analysis for outflows 
following \citet{Das-Chattopadhyay08}. Subsequently, we use the jet sonic point 
properties while obtaining the outflow solution.

\section{solution methodology}

In the accretion process, rotating matter experiences centrifugal repulsion while
falling inward that eventually causes a virtual barrier around the black hole.
Depending on the initial parameters, such a centrifugal barrier can trigger
discontinuous transition in the flow variables in the form of shock. Because of
shock compression, since the post-shock flow ($i.e.,$ PSC) becomes hot and dense,
excess thermal gradient force at PSC is developed that deflects a part of the
infalling matter along the black hole rotation axis in the form of outflows/jets.
In the present analysis, we consider the outflowing matter to be emerged out
between the centrifugal barrier (CB) and funnel wall (FW) surfaces. Following
the above consideration, we move forward to obtain the self-consistent 
accretion-ejection solutions that are coupled via Rankine-Hugoniot (hereafter RH)
shock conditions. In the presence of outflow, the RH conditions are given by,

(i) the conservation of energy flux:
$$
\mathcal{E_{+}} = \mathcal{E_{-}},
\eqno(22a)
$$
where, the quantities with subscripts `$-$' and `$+$' sign indicate the value of the 
variables before and after the shock, respectively.  Here, $\mathcal{E}$ refers
to the local energy of the flow and is obtained from equation (1) as
$\mathcal{E} (x)= v^2(x)/2 + a^2(x)/(\gamma-1) +\Phi_{eff}(x)$.

(ii) the conservation of mass flux:
$$
\dot{M}_{+} = \dot{M}_{-} - \dot{M}_{out}=\dot{M}_{-}(1 - R_{\dot{m}}).
\eqno(22b)
$$
It is already pointed out that during mass loss, a part of the inflowing
matter is emerged out as outflow while the rest is advected in to the
black hole. Here, we denote pre-shock and post-shock  accretion rate
as $\dot{M}_{-}$ and $\dot{M}_{+}$, respectively and define the outflow
rate as $R_{\dot{m}} =\dot{M}_{out}/\dot{M}_{-}$.

\noindent And, finally we have, 

(iii) the conservation of momentum flux:
$$
W_{+} + \Sigma_{+} v_{+}^2 = W_{-} + \Sigma_{-} v_{-}^2.
\eqno(22c)
$$
where, $W$ and $\Sigma$ stands for the vertically integrated pressure and
density stated before  \citep{Das-etal01b}.

In our accretion-ejection model, outflow is considered to be originated from the post shock
region and accordingly, we assume that the outflow is essentially launched
with the same density as in the PSC $i.e., \rho_{j}=\rho_{+}$. Using
equations (3), (20) and (22b), the mass outflow rate is then computed as,
$$
R_{\dot{m}}=\frac{\dot{M}_{out}}{\dot{M}_{-}} 
=\frac{Rv_j(x_s)\mathcal{A}_j(x_s)}{4\pi \sqrt{\frac{1}
{\gamma}} x_s^{3/2}{\Phi_{r}^{\prime}}^{-1/2} a_+ v_-},
\eqno(23)
$$
where, $R$ is the compression ratio defined as $R=\Sigma_+/\Sigma_-$. In addition,
$v_j(x_s)$ and $\mathcal{A}_j(x_s)$ are the jet velocity and the jet area function
calculated at the shock, respectively. Here, we employ the successive iteration method
to calculate $R_{\dot{m}}$ self-consistently. In this method, initially we start with
$R_{\dot{m}} = 0$. Using RH shock conditions, we 
calculate the virtual shock location $x_s^{v}$ for a given set of inflow 
variables, namely $(\mathcal{E}_{inj}, \lambda_{inj}, a_k, \alpha, f)$. Subsequently, we use 
$\mathcal{E}_j = \mathcal{E}_+$ and $\lambda_{j} = \lambda_{+}$ to 
calculate the jet sonic point using equations (19) and (20). Starting form the jet
sonic point, we integrate the jet equations towards the black hole horizon in order to calculate the
jet variables and continue to do so up to the jet base which is equivalently the shock 
location. Utilizing these jet variables, we compute the virtual $R_{\dot{m}}^{v}$ which we
use further in the RH shock conditions to get a new shock location. We continue the iteration
process until the shock location converges to the actual location at $x_s$ and finally obtain 
$R_{\dot{m}}$ corresponding to $x_s$.

\section{Results and Discussions}
 
In this section, we present our model solutions for various sets of initial parameters of the
accretion flow. In particular, here we depict  the effect of dissipation processes, namely
viscosity and cooling on the mass loss from the accretion disc around rotating black hole. 
In the first section, we investigate the effect of viscosity on the outflow rates in the
advection-dominated regime (i.e., $f = 1$) and in the next section, we show 
the effect of cooling on outflow rates in the cooling-dominated regime (i.e., 
$f \rightarrow 0$). As mentioned in \S 2, the pseudo-Kerr potential 
\citep{Chakrabarti-Mondal06} adopted in this work describes the general relativistic
features of space-time geometry quite satisfactorily in the range $-1\leq a_k\leq0.8$.
However, for the sake of completeness,  here we present results beyond $a_k > 0.8$
to examine the overall characteristics of the accretion solution for the maximally
rotating black hole. Though it introduces error $\sim 20\%$
as far as marginally stable orbit is concerned for the maximally rotating case
\citep{Chakrabarti-Mondal06}, we argue that the overall findings of the accretion 
solutions would not deviate severely. 
In addition, it is known that the adiabatic index ($\gamma$) depends on the ratio of
the thermal energy to the rest energy of the flow and the theoretical limit
of $\gamma$ lies in the range $4/3 \leq \gamma \leq 5/3$ \citep{Frank-etal02}. In this
work, for the purpose of representation, we consider $\gamma = 1.4$ all throughout
unless otherwise stated.

\subsection{Effect of viscosity on outflow rates}

\begin{figure}
\begin{center}
\includegraphics[width=0.45\textwidth]{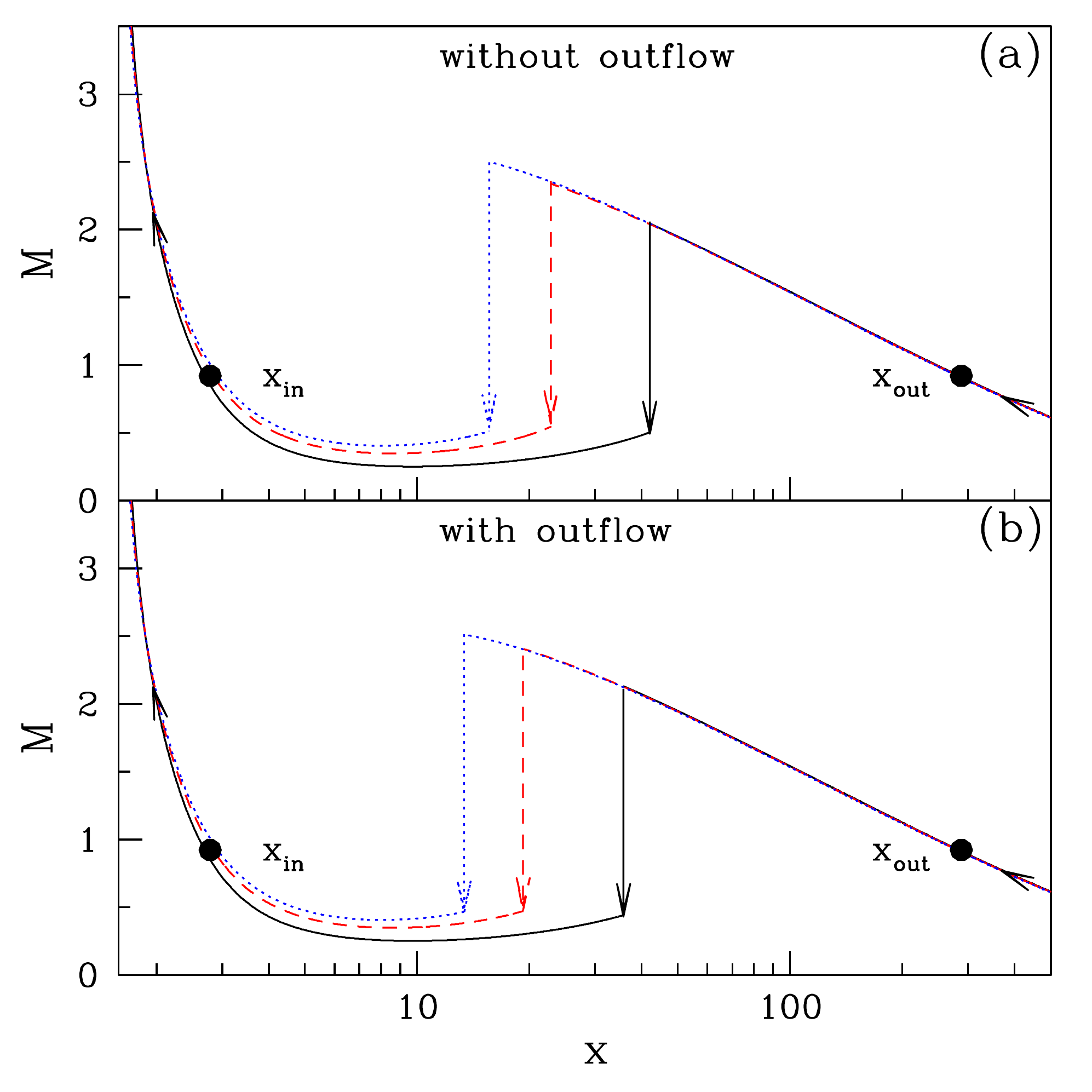}
\end{center}
\caption{
Variation of inflow Mach number $M=(v/a)$ with radial distance 
$(x)$. Solid, dashed and dotted curves are for $\alpha $ = 0.0 (black),
0.003 (red) and 0.0046 (blue), respectively. Upper panel is for without
outflow and the lower panel is for with outflow. Here, we fix 
$a_k = 0.76$. See text for details.
}
\end{figure}

In Fig. 2, we present the global dissipative viscous accretion solutions
around the rotating black holes for various sets of inflow parameters.
All these solutions contain shock waves. Here, the flow parameters
at the outer edge is chosen as $x_{\rm inj} = 500$, $\lambda_{\rm inj}=2.656$ and 
$\mathcal{E}_{\rm inj}=0.00135$, respectively and the black hole spin is considered
as $a_k = 0.76$. In Fig. 2a, we show the Mach number $(M=v/a)$ variation of the
inflowing matter with the radial distance in absence of mass loss. The solid, dashed
and dotted curves are for the viscosity parameter $\alpha =0.0$ (black), $\alpha =0.003$ (red) and
$\alpha =0.0046$ (blue), respectively. We find that the shock front moves inward
towards the black hole horizon with the increase of  $\alpha$. This is simply because the
increase of  $\alpha$ enhances the angular momentum transport towards the
outer edge of the disc that eventually instigates the debilitating of centrifugal
barrier around the black hole. Effectively, this compels the shock front to move inward
in order to preserve the pressure balance across it. In the figure, vertical arrows
indicate the shock transition and the location of the shock are obtained as
$x_s = 42.03$, $22.85$ and $15.59$ for inviscid case, $\alpha =0.0$, $0.003$ and $0.0046$, 
respectively. In Fig. 2b, we display the accretion solution in presence of mass
loss for the same set of inflow parameters as in Fig. 2a. In the present model,
the outflow is launched from the post-shock 
region. Therefore, when a part of the inflowing matter is emerged out 
from the disc as outflow, the effective pressure at the inner part of the
disc is reduced and eventually the shock front is propelled towards the black hole 
even further in order to maintain the pressure balance across the shock.
The solid, dashed and dotted curves denote the results corresponding to
the viscosity parameters, shock locations and outflow rates as,
$(\alpha, x_s, R_{\dot{m}}) = (0.0, 35.76, 0.0346)$ (black), $(0.003, 19.24, 0.0220)$ (red) and 
$(0.0046, 13.36, 0.0182)$ (blue), respectively. In both the panels, filled circles indicate
the inner $(x_{\rm in})$ and outer $(x_{\rm out})$ sonic points and arrows represent
the flow direction.

\begin{figure}
\begin{center}
\includegraphics[width=0.45\textwidth]{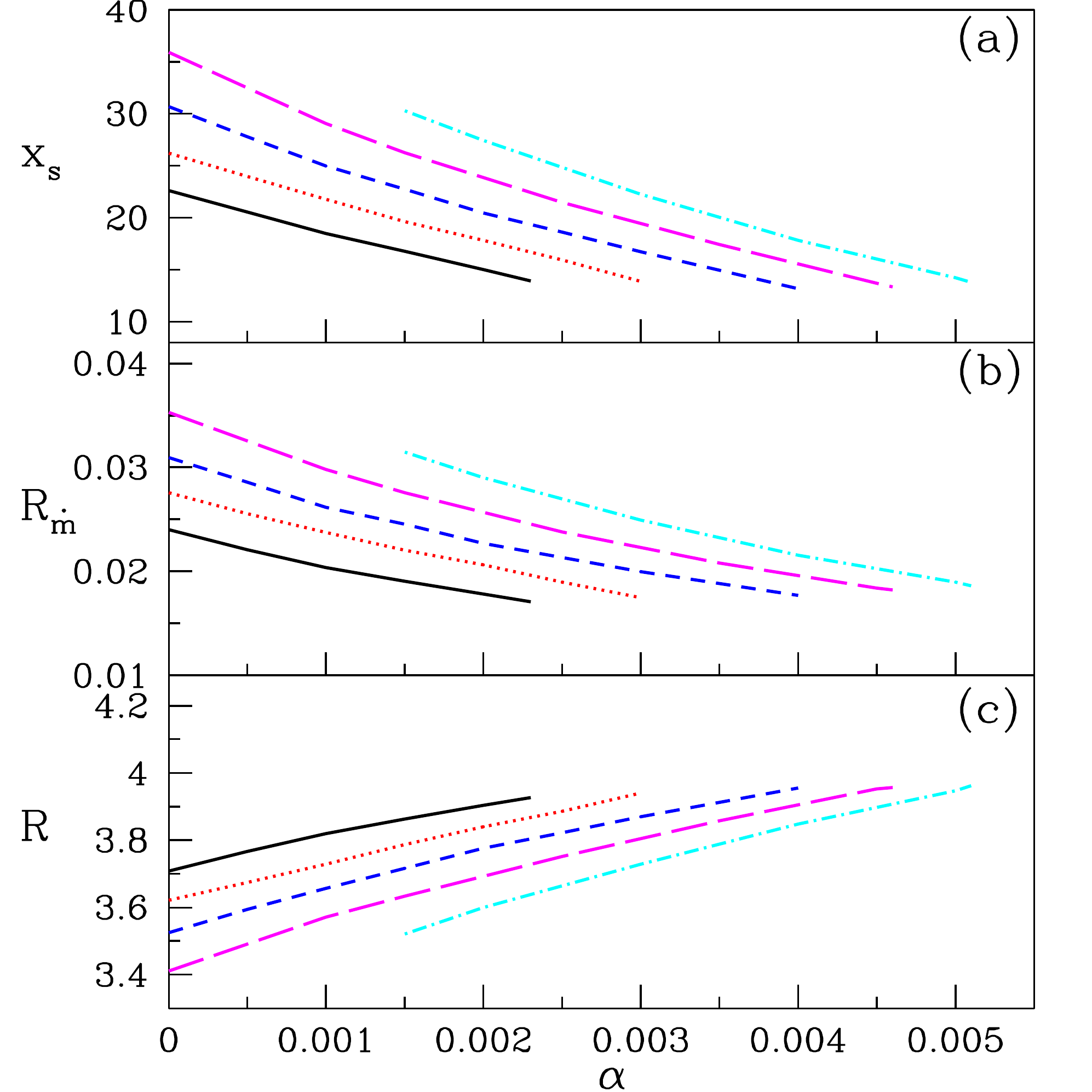}
\end{center}
\caption{
Variation of (a) shock location $(x_s)$, (b) outflow rate 
$(R_{\dot{m}})$ and (c) compression ratio $(R)$ as a function of viscosity 
parameter $\alpha$ for $a_k = 0.78$ to $0.70$, where $\Delta a_k$ = 0.02 
(right to left). Here, we fix the outer boundary at  $x_{\rm inj} = 500$. The angular
momentum and energy of the flow at $x_{\rm inj}$ is chosen as $\lambda_{\rm inj} = 2.656$
and $\mathcal{E}_{\rm inj} = 0.00135$, respectively. See text for further details.
}
\end{figure}

In Fig. 3, we investigate the effect of viscosity parameter on the inflow-outflow
solutions around the rotating black holes. Here, we fix the injection radius of the inflowing
matter at $x_{\rm inj} = 500$ and the corresponding inflow energy ($\mathcal{E}_{\rm inj}$)
and angular momentum ($\lambda_{\rm inj}$) at $x_{\rm inj}$ are 
$(\mathcal{E}_{\rm inj}, \lambda_{\rm inj}) = (0.00135, 2.656)$, respectively. We vary the spin of
the black hole $a_k$ from $0.78$ to $0.7$ from the right most (dot-big-dashed) to the
left most curves (solid) with an
interval $\Delta a_k$ = 0.02. In the upper panel (Fig. 3a), we show the variation of 
shock location with the viscosity parameter $\alpha$ in presence of mass loss.
As it is seen in Fig. 2, here also we find that shock ($x_s$) proceeds closer to the
black hole with the increase of viscosity parameter ($\alpha$). This clearly indicates that
the effective size of the PSC decreases with the increase of $\alpha$.
As a consequence, the fraction of inflowing matter intercepted by the PSC
is reduced that eventually produce feeble outflow rate when $\alpha$ is 
increased. In Fig. 3b, we plot the outflow rates ($R_{\dot{m}}$) corresponding to
Fig. 3a. In Fig. 3c, we study the compression ratio 
($R$) of shocked accretion flow in
presence of mass loss as function of $\alpha$. Compression ratio essentially
measures the density compression across the shock and is defined as the ratio
of post-shock density to the pre-shock density.
As the shock moves closer to the horizon, flow experiences enhanced
compression and therefore, $R$ increases with the increase of $\alpha$. 

\begin{figure}
\begin{center}
\includegraphics[width=0.45\textwidth]{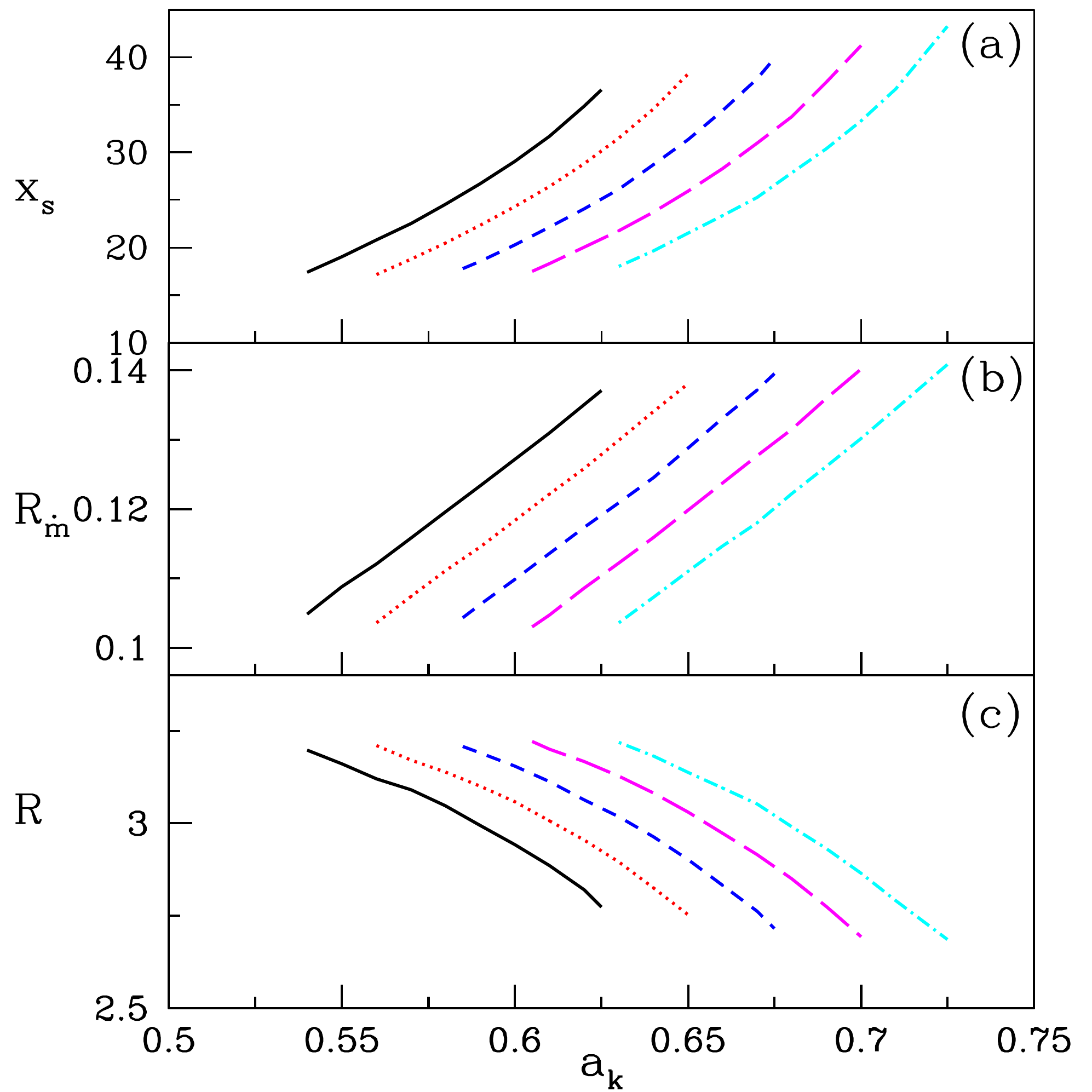}
\end{center}
\caption{
Plot of (a) shock location $(x_s)$, (b) outflow rate $(R_{\dot{m}})$ and
(c) compression ratio $(R)$ variation as function of spin $a_k$ 
for $\alpha = 0.003$ to $0.007$, where $\Delta \alpha$ = 0.001 (left to 
right). Here, we fix the outer boundary at $x_{\rm inj} = 300$ and choose
$\lambda_{\rm inj} = 2.80$ and $\mathcal{E}_{\rm inj} = 0.0025$, respectively.
See text for details.
}
\end{figure}

In Fig. 4, we show the effect of black hole rotation on the shock properties and the 
mass outflow rate. For this analysis, we choose the outer boundary of the disc
at $x_{\rm inj} = 300$ with specific energy and angular momentum as
$\mathcal{E}_{\rm inj} = 0.0025$ and $\lambda_{\rm inj} = 2.80$, respectively.
In the upper $(a)$, middle $(b)$ and bottom $(c)$ panels of Fig. 4, we
plot shock locations $(x_s)$, outflow rates $(R_{\dot{m}})$ and
compression ratio $(R)$ as function of black hole spin $(a_k)$, respectively.
In all the panels, we vary the viscosity parameter from $\alpha = 0.003$
to $0.007$ with an interval $\Delta \alpha = 0.001$ from left most to
the right most curve. We observe that {\it for fixed outer boundary
condition}, the shock location ($x_s$) moves away from the black hole horizon
with the increase of black hole rotation $a_k$ for flows with fixed $\alpha$.
As a result, the effective area of PSC intercepted by the inflowing
matter is increased and consequently, the amount of the inflowing
matter deflected by the PSC is also increased yielding enhanced mass
outflow rates $R_{\dot{m}}$. In contrary, we find that $R_{\dot{m}}$
decreases with the increase of $\alpha$ for fixed $a_k$ as seen in
Fig. 3.  In addition, as $x_s$ recedes away from the black hole horizon with
the increase of $a_k$ for flows with constant $\alpha$, the compression
ratio $(R)$ is also decreased as shown in Fig. 4c. On the other hand, for fixed
$a_k$, as $\alpha$ increased, both $x_s$ and $R_{\dot m}$ decreases
whereas $R$ increases. This findings
are compatible with results of \citet{Aktar-etal15}.

\begin{figure*}
\begin{center}
\includegraphics[angle=00,width=0.45\textwidth]{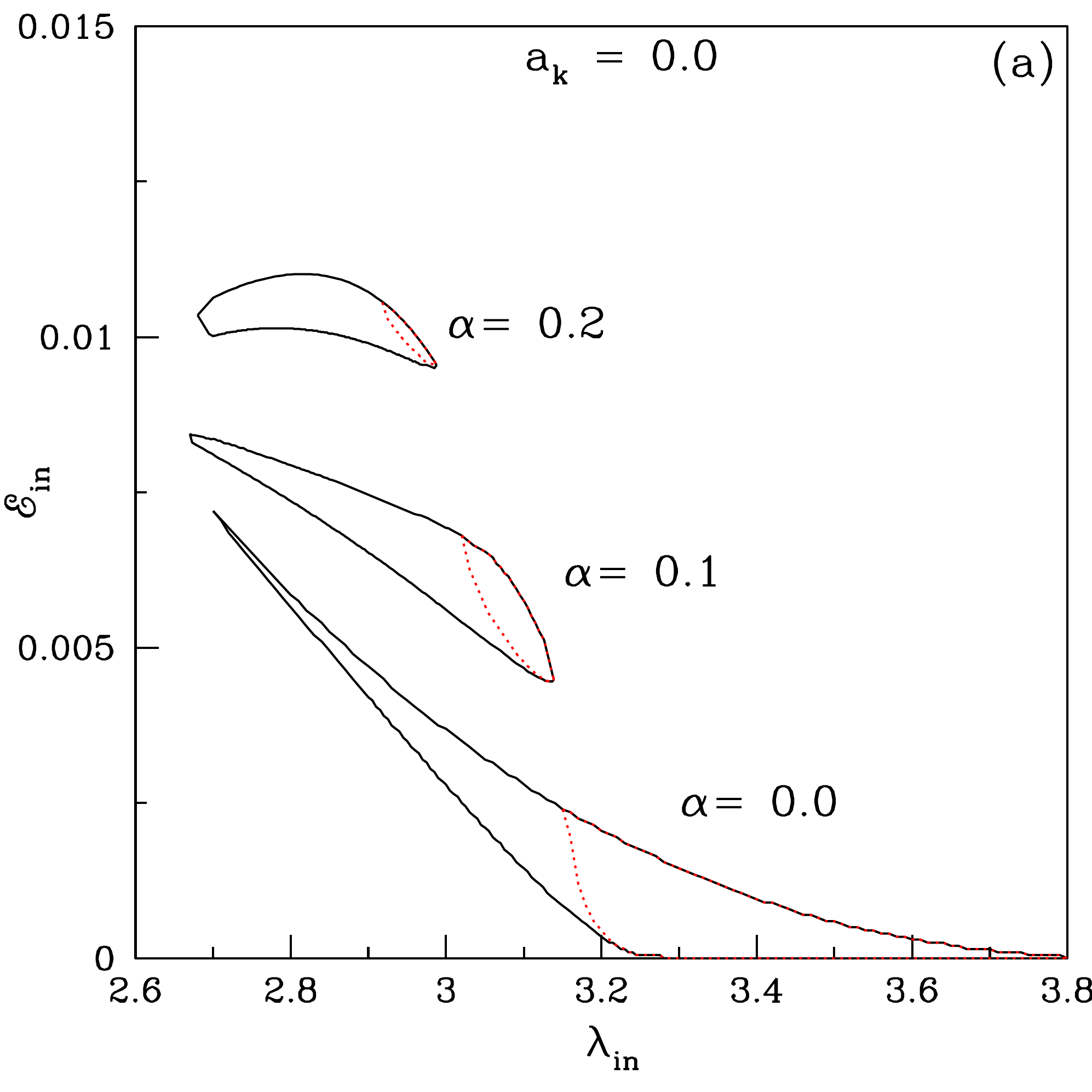}
\includegraphics[angle=00,width=0.45\textwidth]{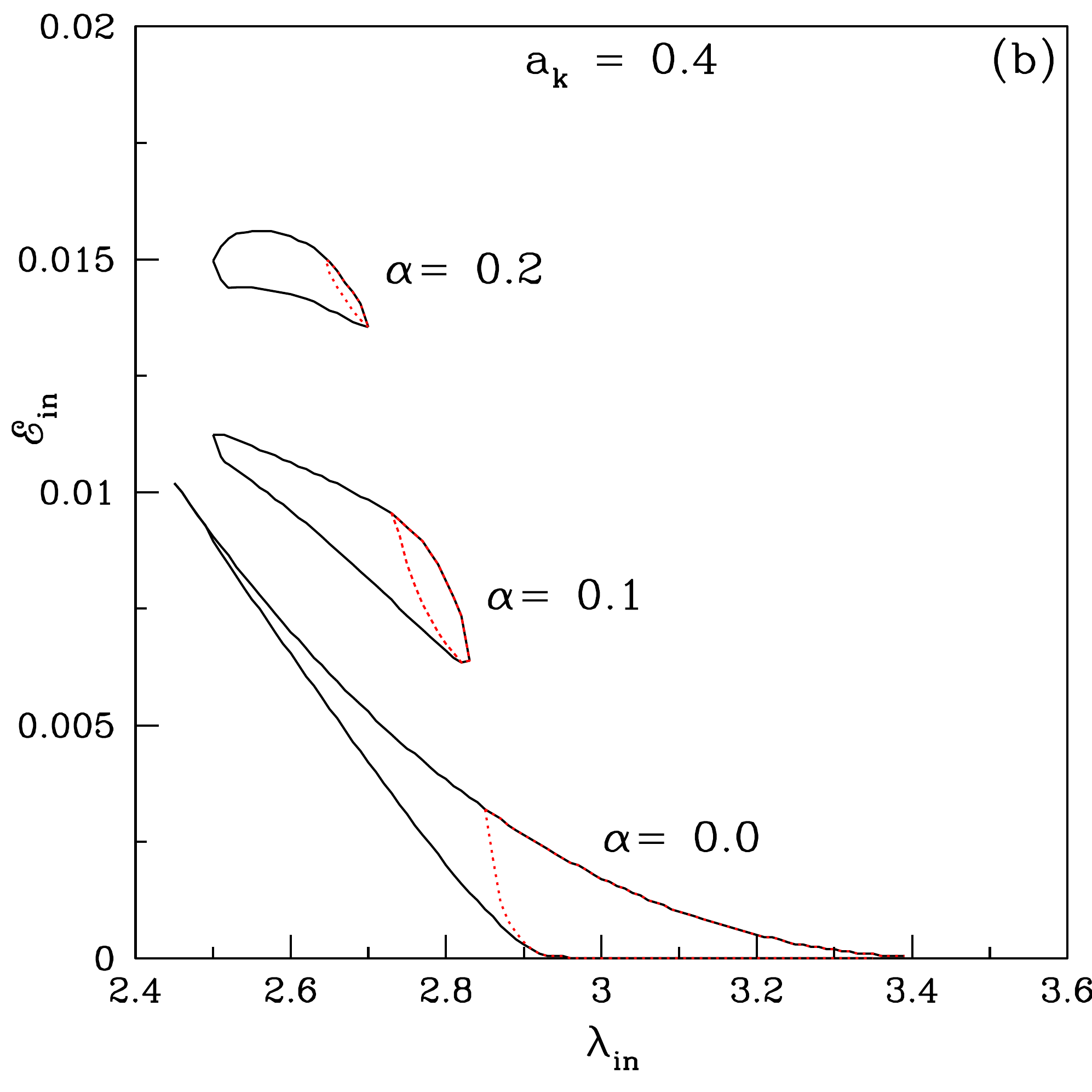}
\includegraphics[angle=00,width=0.45\textwidth]{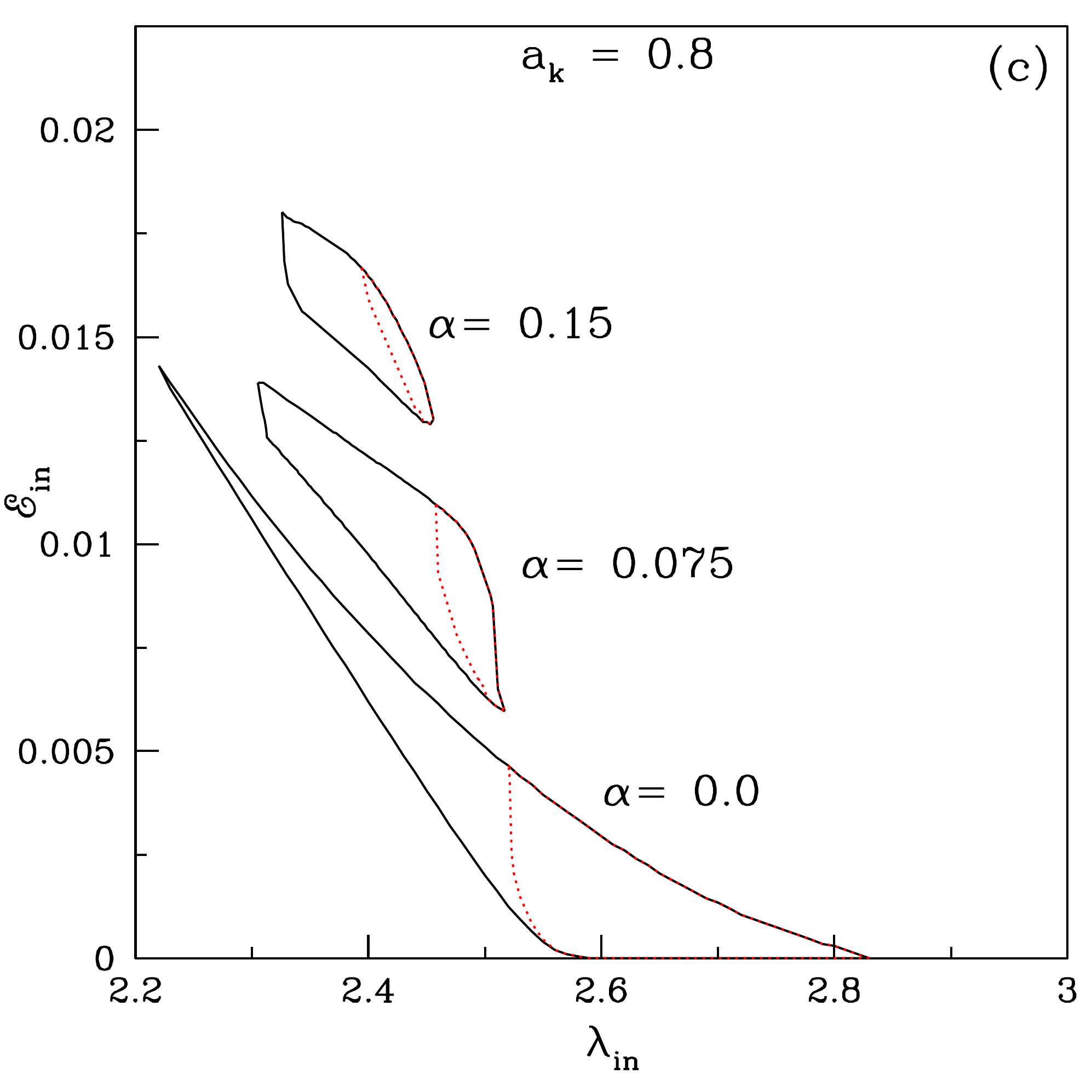}
\end{center}
\caption{
Identification of parameter space for shock in $\lambda_{\rm in} - {\cal E}_{\rm in}$
plane is shown as function of viscosity parameter ($\alpha$) for various black
hole spin parameter ($a_k$). Solid curves denote the shock parameter space in
absence of mass loss whereas dotted curves separate the shock parameter 
space in presence of mass loss. Fig. 5a, 5b and 5c illustrate the results
for $a_k = 0.0$, $0.4$ and $0.8$, respectively. See text for further details.
}
\end{figure*}

We proceed further to examine the region of parameter space that permits
stationary shock solutions. As it is already pointed out (see Fig. 2-4)
that the shock induced global accretion solutions are not the isolated solutions,
instead the solutions of this kind can be obtained for a wide range of inflow
parameters. Moreover, in our model, since the outflow is originated from the
PSC, it is worthy to identify the shock parameter space in terms of the viscosity
parameter ($\alpha$) and the spin of the black hole ($a_k$). In Fig. 5, we classify
the region of the shock parameter space spanned by the angular momentum
($\lambda_{\rm in}$) and energy ($\mathcal{E}_{\rm in}$) of the flow measured at 
the inner sonic point ($x_{\rm in}$). The results displayed in Fig. 5(a-c) are obtained
for $a_k =0.0$ ($a$), $0.4$ ($b$) and $0.8$ ($c$), respectively. The corresponding 
viscosity parameters are marked in each panel. The solid curves separate
the shock parameter space when mass loss from the disc is ignored whereas the dotted
curves represent the shocked parameter space including mass loss. In presence of 
outflow, the effective area of the PSC becomes smaller (see Fig. 2) and therefore, the resulting
parameter space is reduced in comparison to the no mass loss case, particularly in the
lower angular momentum and higher energy side. When the input parameters
are chosen from these parts of the parameter space ignoring outflow, accretion
flow encounters shock transition very close to the black hole \citep{Das-etal01a}.
After that when outflow is allowed to be launched from the PSC, shock ceases to exist.
Subsequently, we study the role of viscosity in classifying the parameter space.
In the accretion flow, the presence of viscosity manifests dual effects. Firstly, viscosity
transports angular momentum outward reducing its value at the inner edge of the
disc and secondly, flow is heated up due to the effect of viscous dissipation. Hence, 
as the viscosity is increased, the shock parameter space is shrunk from the both ends
of the angular momentum range and also shifted towards the higher energy domain.
Moreover, with the inclusion of viscosity, flow becomes dissipative and consequently,
the possibility of standing shock formation is reduced \citep{Chakrabarti-Das04,Das07}
and finally parameter space for standing shock disappears when
the critical viscosity limit is reached. 
Interestingly, time varying shock solutions may still exist beyond the critical
viscosity limit \citep{Chakrabarti-Das04,Das-etal14}. Numerical simulations already explored
these oscillatory behavior of shock solutions
\citep{Lee-etal11,Das-etal14,Das-Aktar15,Sukova-Jianuk15} which 
successfully describe the Quasi-periodic Oscillations (QPOs) and periodic
mass loss phenomena observed in several astrophysical black hole systems
\citep{Chakrabarti-etal02,Nandi-etal12,Radhika-etal16a}.
A comprehensive study of time dependent global accretion solution around
rotating black hole is considered as a future work as it is beyond the scope
of the present paper and will be reported elsewhere.

\begin{figure}
\begin{center}
\includegraphics[width=0.45\textwidth]{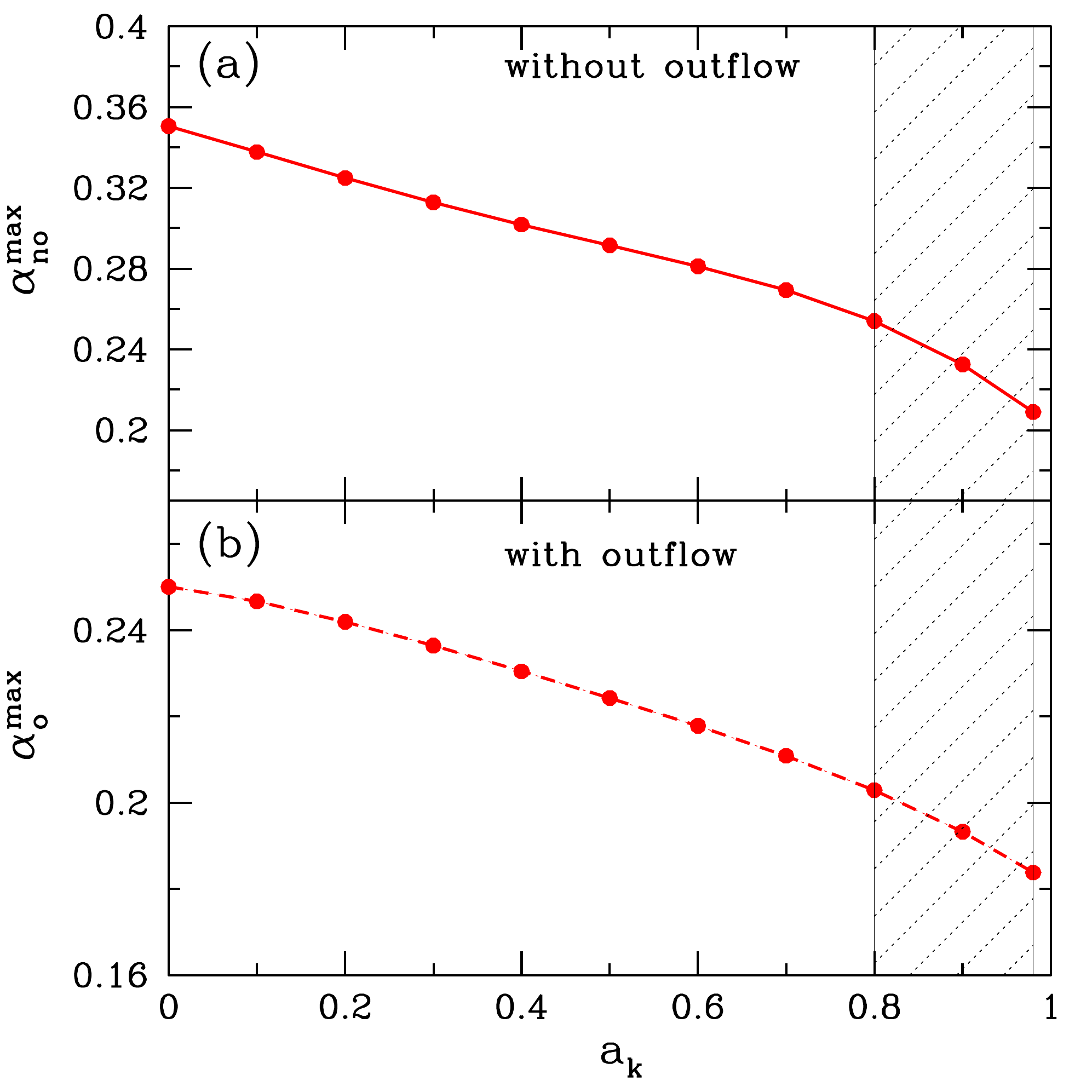}
\end{center}
\caption{
Variation of maximum viscosity parameter $(\alpha^{\rm max})$ with 
spin of the black hole $a_k$ that allows standing shock. In the
upper panel $(a)$, filled circles connected with the solid curve
denotes $\alpha_{\rm no}^{\rm max}$ in absence of outflow and in the lower panel
$(b)$, filled circles joined with the dashed curve represents 
$\alpha_{\rm o}^{\rm max}$ in presence of outflow. In both the panels, results within
the shaded region are obtained for $a_k > 0.8$. See text for
further details.
}
\end{figure}

We continue the study of shock properties in terms of the input flow parameters. Towards this,
we investigate the maximum value of the viscosity parameters ($\alpha^{\rm max}_{\rm no}$ and
$\alpha^{\rm max}_{\rm o}$)
that allows steady shock solutions in absence ($R_{\dot{m}} = 0$) 
as well as in presence ($R_{\dot{m}} \ne 0$) of mass loss.
While doing so, we first fix the black hole spin
($a_k$) and vary the remaining flow parameters freely. In Fig. 6,
we depict the variation of $\alpha^{\rm max}$ as function of $a_k$ where the
upper panel is for without mass loss case and the lower panel is for with
mass loss case. We already pointed out that the pseudo-Kerr potential adopted here describes
the space time geometry around the black hole satisfactorily for $a_k \le 0.8$. However, 
we present results for $a_k>0.8$ which are depicted with shaded region. Here,
we argue that the obtained results demonstrate the overall findings
of the accretion solutions at least qualitatively \citep{Chakrabarti-Mondal06}.
We find that the upper limit of viscosity parameter $\alpha^{\rm max}$ for
shock is anti-correlated with the spin $a_k$ in both the panels. In general,
at the lower viscosity parameter,
the sub-Keplerian flow joins with the Keplerian disc far away from the black hole horizon. This
eventually increases the possibility of possessing multiple sonic points including shock waves.
On the other hand, at the high viscosity limit, ($i. e.$, $\alpha > \alpha^{\rm max}$),
Keplerian disc comes very close to the black hole horizon allowing the flow to pass
only through the single sonic point  \citep{Chakrabarti96a}. In addition, the rotation of the
black hole drags the accreting matter towards the horizon by its strong gravitational pull.   
Therefore, the possibility of the formation of steady shock in the higher viscosity
domain for increasing $a_k$ becomes rarer.
It is to be noted that the obtained $\alpha^{\rm max}$ for non-rotating black hole is
consistent with the results of \citet{Chakrabarti-Das04}. 
When outflow is considered, the possibility of shock formation is decreased 
(see Fig. 5) and therefore, the obtained $\alpha ^{\rm max}$ for $R_{\dot{m}} \ne 0$
becomes smaller compared to case with $R_{\dot{m}} = 0$ \citep{Kumar-Chattopadhyay13}.

\begin{figure*}
\begin{center}
\includegraphics[angle=00,width=0.425\textwidth]{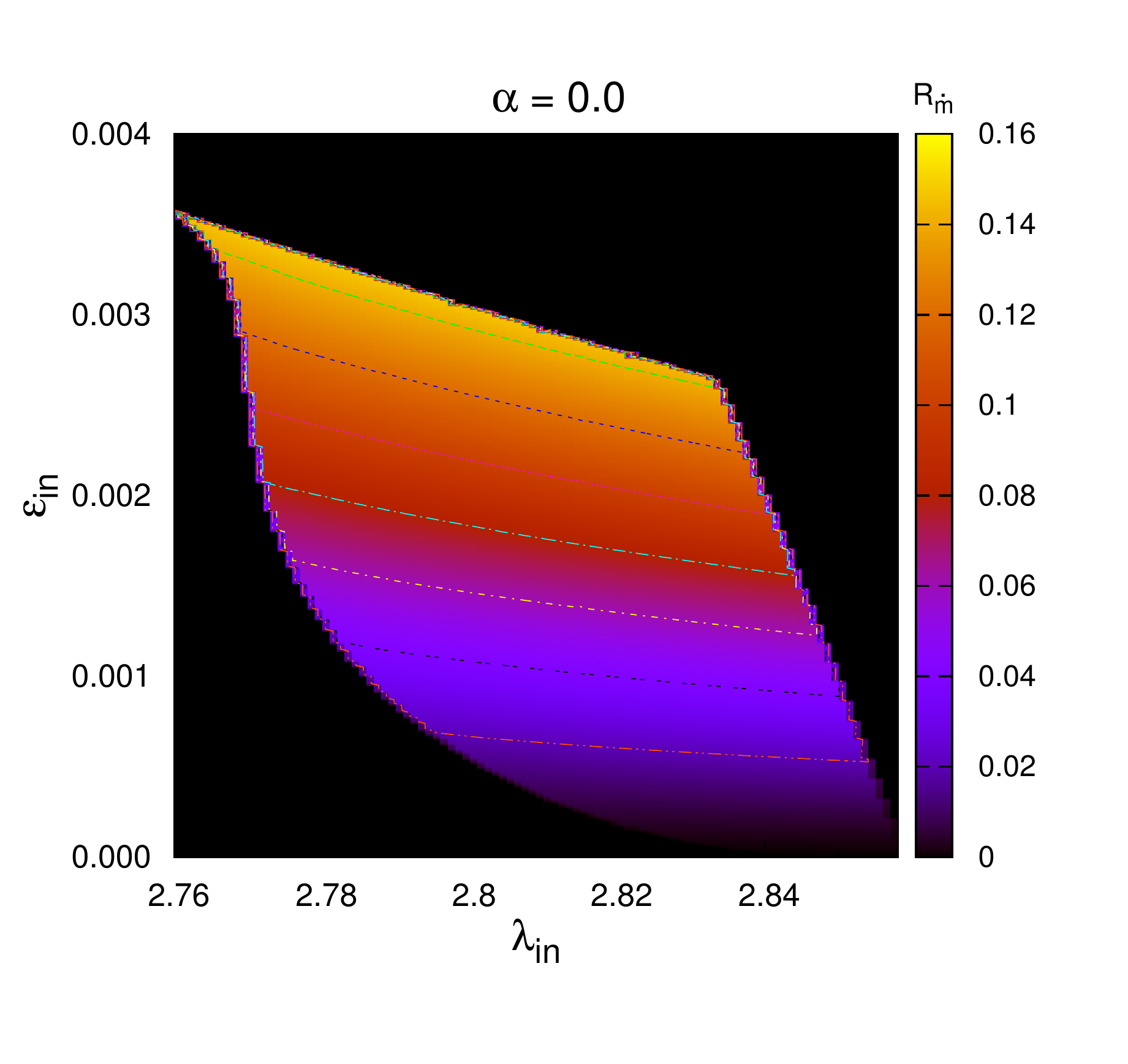}
\includegraphics[angle=00,width=0.425\textwidth]{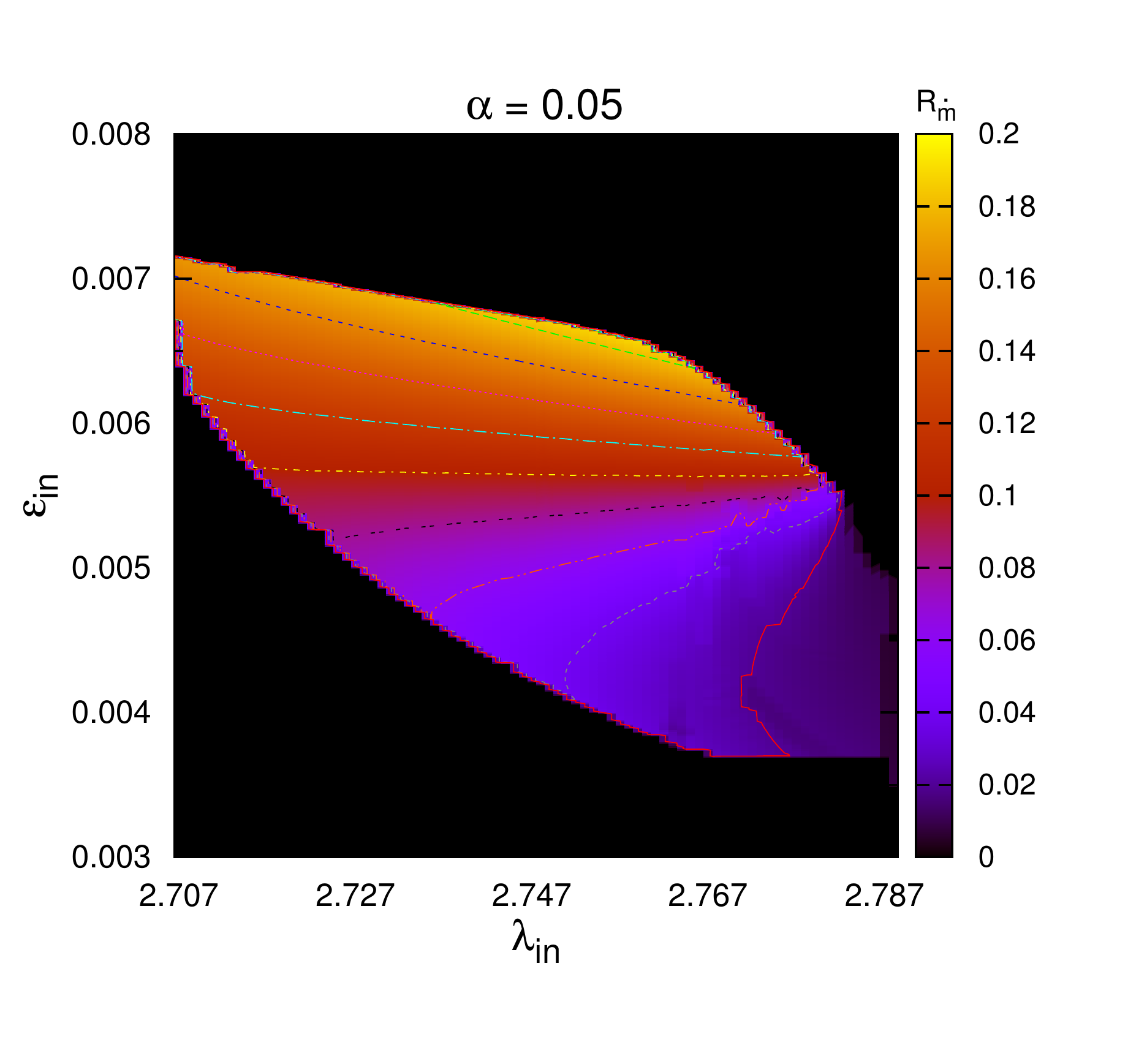}
\includegraphics[angle=00,width=0.425\textwidth]{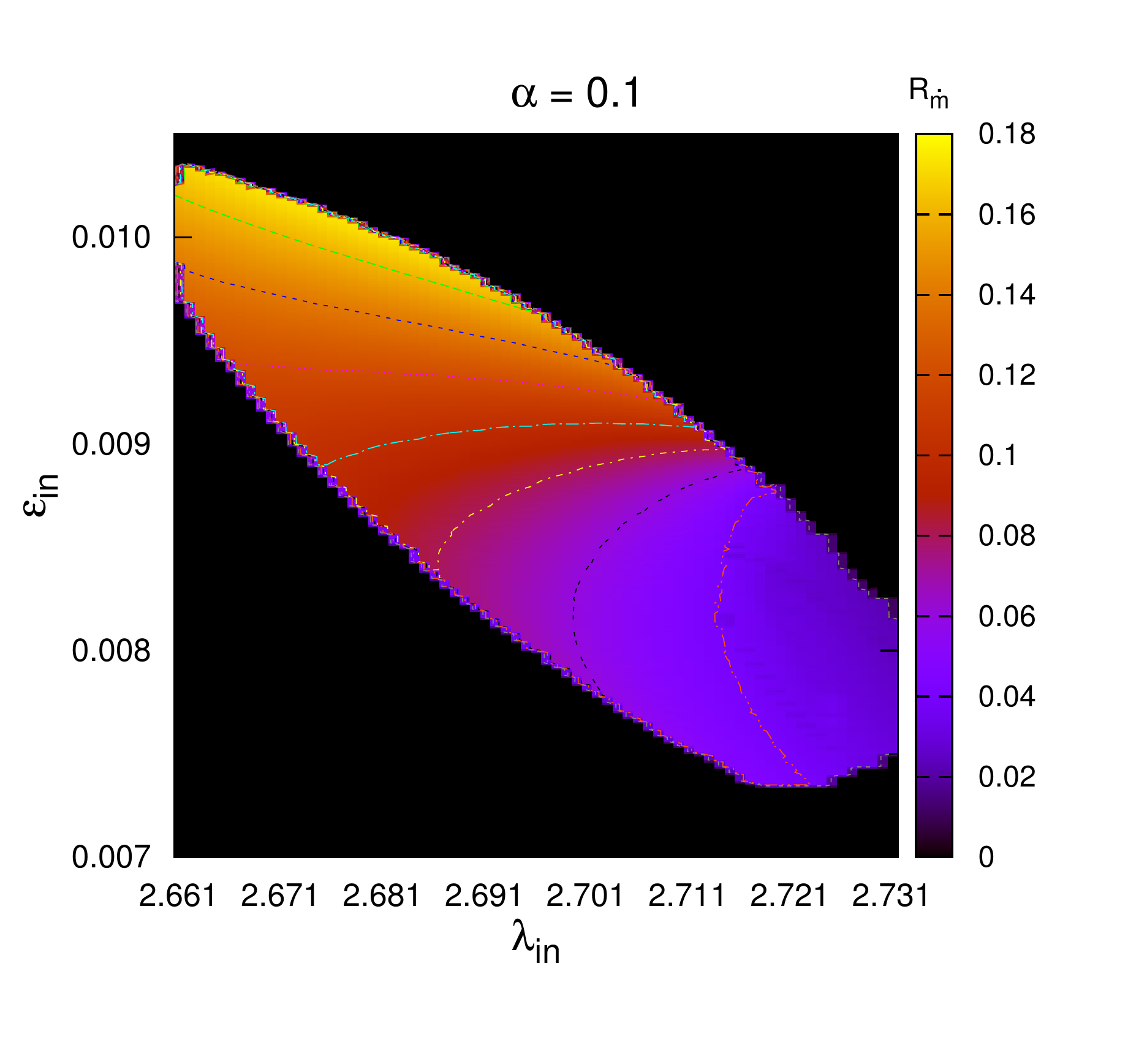}
\includegraphics[angle=00,width=0.425\textwidth]{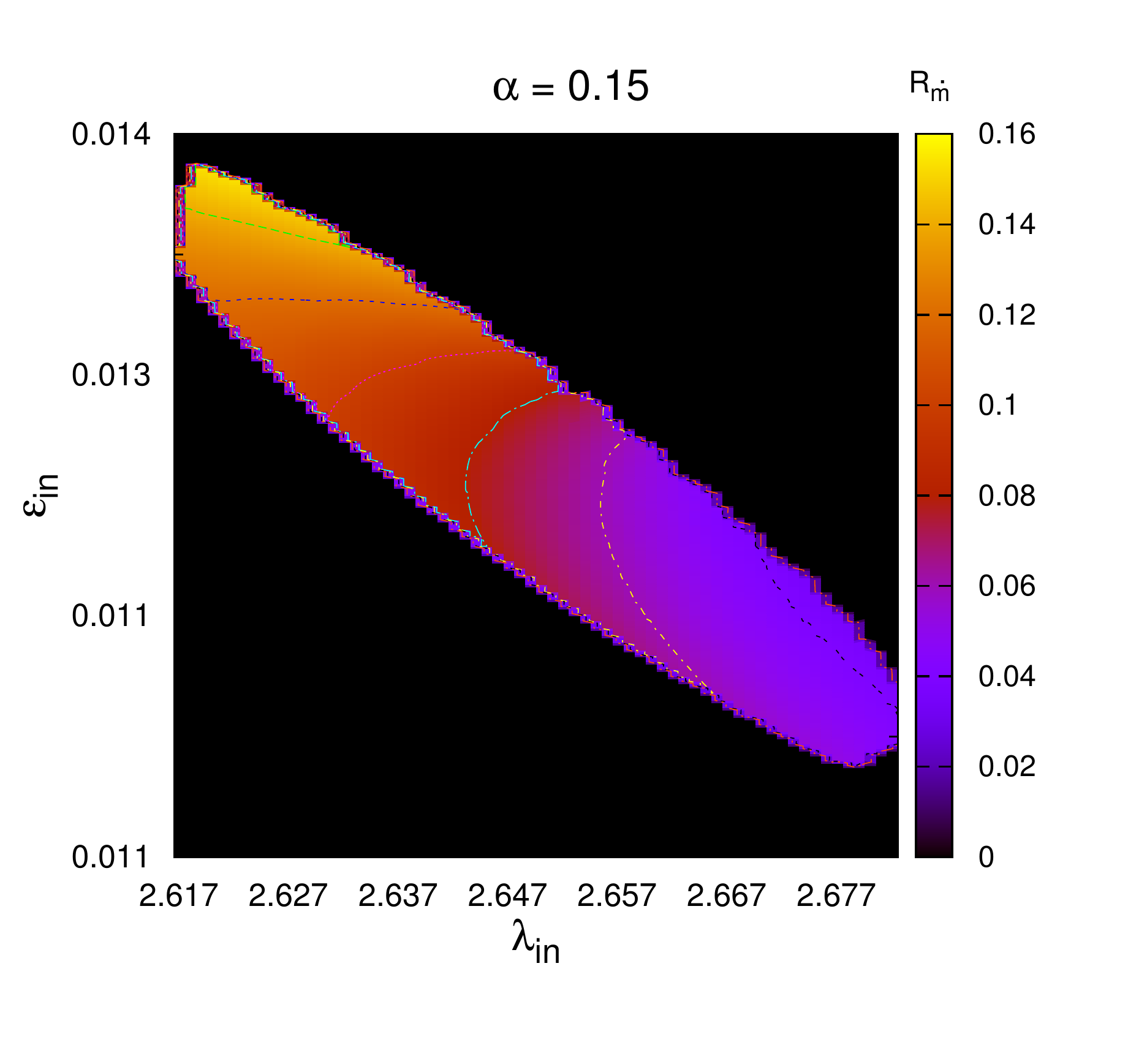}
\end{center}
\caption{Two dimensional (2D) surface projection of three dimensional (3D) 
plot of flow energy ($\mathcal{E}_{\rm in}$), angular momentum 
($\lambda_{\rm in}$) and outflow rates ($R_{\dot{m}}$). In each panel, vertical
color coded bar represents the estimated range of outflow rates. Here, the viscosity 
parameters are chosen as $\alpha$ = 0.0, 0.05, 0.1 and 0.15, respectively and
we fix black hole spin as $a_k$ = 0.5. See text for details. 
}
\end{figure*}

In Fig. 7, we present the outflow parameter spaces for various $\alpha$ where each
panel demonstrates the two dimensional surface projection of the three dimensional
plot spanned with $\mathcal{E}_{\rm in}$, $\lambda_{\rm in}$ and $R_{\dot{m}}$, respectively.
Here, we choose $a_k = 0.5$ and vary the viscosity parameter as $\alpha = 0.0$ (top-left),
$0.05$ (top-right), $0.1$ (bottom-left) and $0.15$ (bottom-right).
In each panel, the color-coded bar represents the range of $R_{\dot{m}}$ which is
obtained from our model calculation. Moreover, the color coded  contours are drawn
with an interval of $\Delta R_{\dot{m}}$ = 0.02 starting from $R_{\dot{m}}^{\rm max}$
up to its minimum value. In other words,  the different colors embedded with color coded 
contours illustrate the overall span of  $2\%$ outflow rate altogether. In addition,
we find that the effective bounded region of the outflow parameter space gradually
shrinks with the increase of the viscosity parameter $\alpha$ as is seen in Fig. 5. 
It is to be noted  that in order to obtain the maximum outflow rate ($R_{\dot{m}}^{\rm max}$),
the inflow parameters must lie in the higher $\mathcal{E}_{\rm in}$ and
lower $\lambda_{\rm in}$ domains irrespective to the value of the viscosity parameter $\alpha$.

\begin{figure}
\begin{center}
\includegraphics[width=0.45\textwidth]{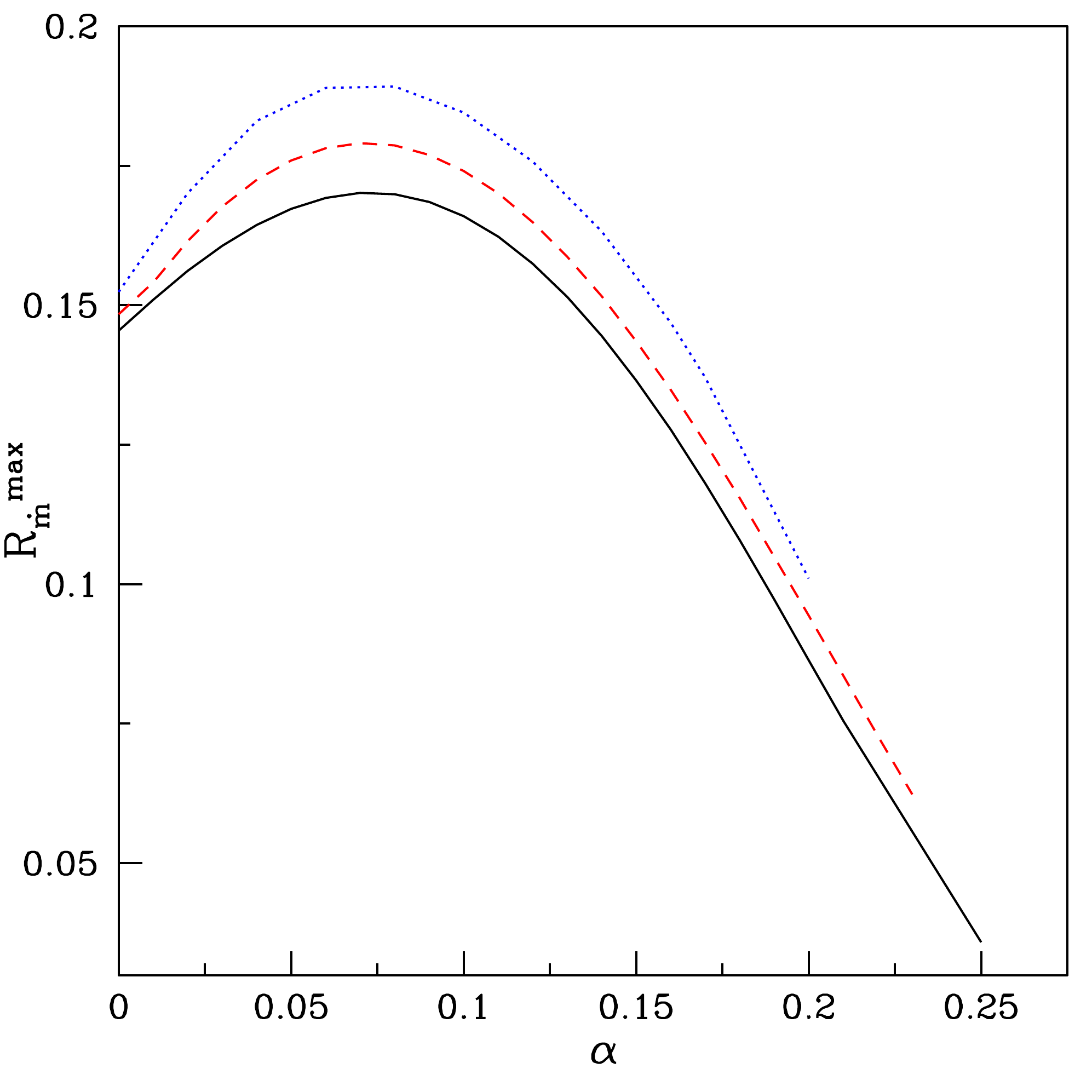}
\end{center}
\caption{General behaviour of maximum outflow rates 
$R_{\dot{m}}^{\rm max}$ as function of viscosity parameter 
$\alpha$.  Solid, dashed and dotted curves are for $a_k = 0.0$ (black),
$0.4$ (red) and $0.8$ (blue), respectively. See text for details.
}
\end{figure}

In the course of our study, we put an effort to compute the maximum outflow
rate ($R_{\dot{m}}^{\rm max}$) that is being originated from the disc. While doing so,
for a given $a_k$, we calculate $R_{\dot{m}}^{\rm max}$ in terms of viscosity
parameter ($\alpha$) by varying the remaining inflow parameters independently. 
The obtained result is depicted in Fig. 8 where we show the  variation of maximum
outflow rates $R_{\dot{m}}^{\rm max}$ with $\alpha$. In the figure, solid, dashed and dotted
curves denote the results corresponding to $a_k = 0.0$ (black), $0.4$ (red)
and $0.8$ (blue), respectively. We find that $R_{\dot{m}}^{\rm max}$ initially increases gradually
with $\alpha$ and reaches to its maximum value and finally starts decreasing with the
increasing $\alpha$ as shown in the figure. We observe that for a given $\alpha$,
$R_{\dot{m}}^{\rm max}$ weakly correlates with $a_k$ which seems to be consistent 
with the results of inviscid flow as reported in \citet{Aktar-etal15}. Overall, 
we find that the maximum outflow rate typically lies in the range
$3\% \lesssim R_{\dot{m}}^{\rm max} \lesssim 19\%$ for widespread viscosity
parameter ($\alpha$).

\subsection{Effect of cooling on outflow rates}

\begin{figure}
\begin{center}
\includegraphics[width=0.45\textwidth]{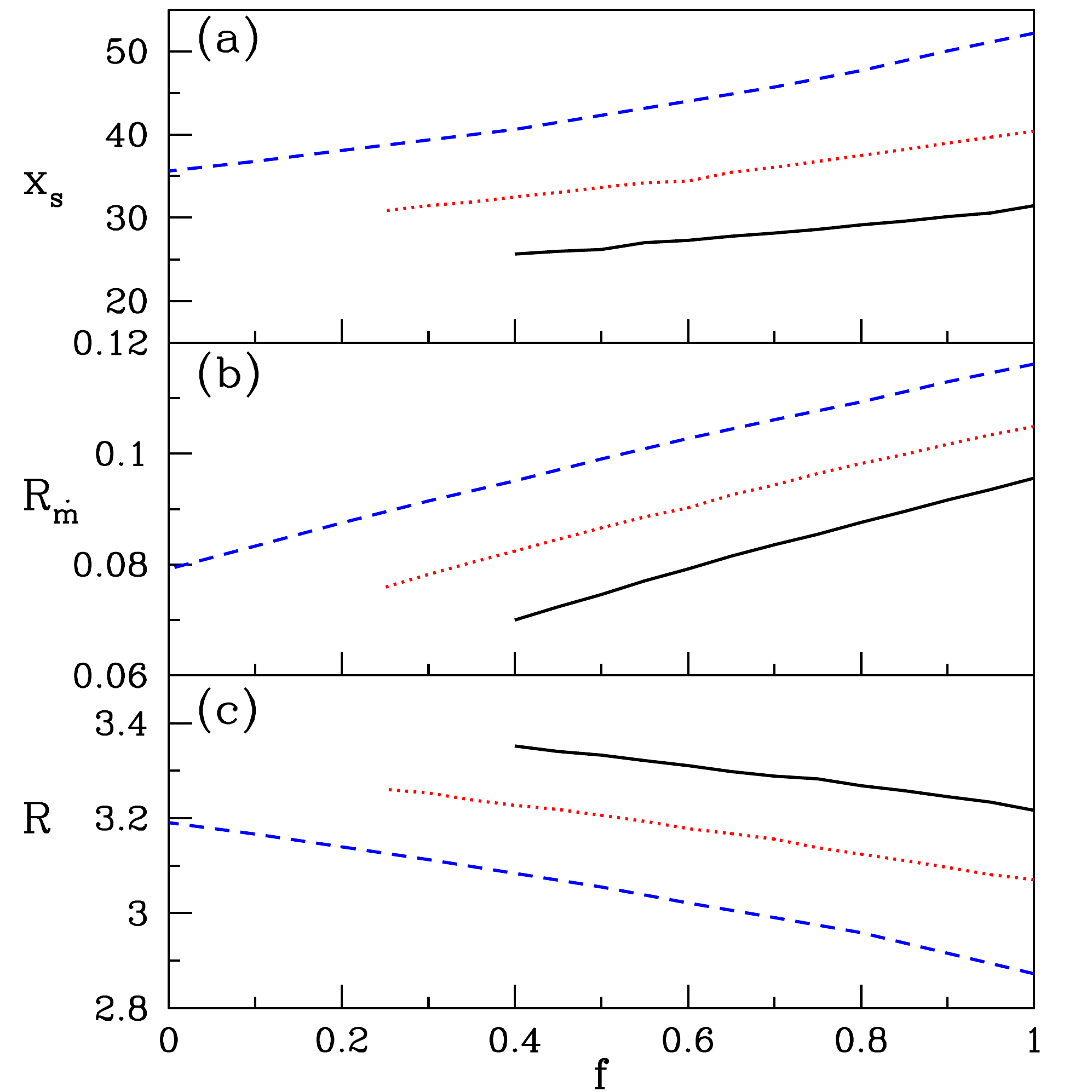}
\end{center}
\caption{
Plot of shock locations ($x_s$) (upper panel), outflow rates ($R_{\dot{m}}$)
(middle panel) and compression ratio ($R$) (lower panel) as function of the
cooling factor $f$. Solid, dotted and dashed curves denote the results for
$\lambda_{\rm inj}$ = 3.975 (black), 4.0 (red) and 4.025 (blue), respectively.
Here, we fix the outer boundary at $x_{\rm inj} = 500$ and flow energy as $\mathcal{E}_{\rm inj} 
=0.0015$.  The viscosity parameter and spin values are chosen as $\alpha = 0.05$ and
$a_k = 0.5$, respectively.
}
\end{figure}
Until now, we present all the result in the advection dominated 
regime i.e., $f = 1$. In this section, we discuss the effect of 
cooling on outflow rates around rotating black hole. Here, we consider 
the parametric cooling as prescribed by \citet{Narayan-Yi94} 
for our analysis. In Fig. 9, we plot the variation of shock locations
$(x_s)$ (upper panel), outflow rates $(R_{\dot{m}})$ (middle panel) 
and compression ratio ($R$) (lower panel) with cooling factor $f$.
To obtain the result, here we choose $x_{\rm inj} = 500$,
$\mathcal{E}_{\rm inj} = 0.0015$, $\alpha = 0.05$ and $a_k = 0.5$,
respectively. In the figure, solid, dotted and dashed curves represent
the results corresponding to the angular momentum at $x_{\rm inj}$ as
$\lambda_{\rm inj} = 3.975$, $4.0$ and $4.025$, respectively. 
In Fig. 9a, we find that the shock location proceeds towards 
the black hole horizon with the decrease of cooling parameter ($f$).
Actually, in presence of cooling, flow loses its energy while accreting 
towards the black hole. In particular, cooling is more effective 
in the post-shock region compared to the pre-shock region due to the 
enhanced density and temperature distributions. This reduces the
post-shock thermal pressure which eventually compels the shock front
to move towards the black hole in order to maintain pressure
balance across the shock. As a consequence, the size of the PSC is
decreased with the increase of cooling causing the reduction of $R_{\dot{m}}$
as depicted in Fig. 9b. In addition, as the shock front moves closer to
the black hole, the corresponding compression  ratio ($R$) is enhanced
with cooling as shown in Fig. 9c. We also observe that for a fixed cooling
parameter, the shock location recedes away from black hole with the
increase of angular momentum at the outer edge ($x_{\rm inj}$).
This provides an indication that the shock transition seems to be
centrifugally driven.

\begin{figure}
\begin{center}
\includegraphics[width=0.45\textwidth]{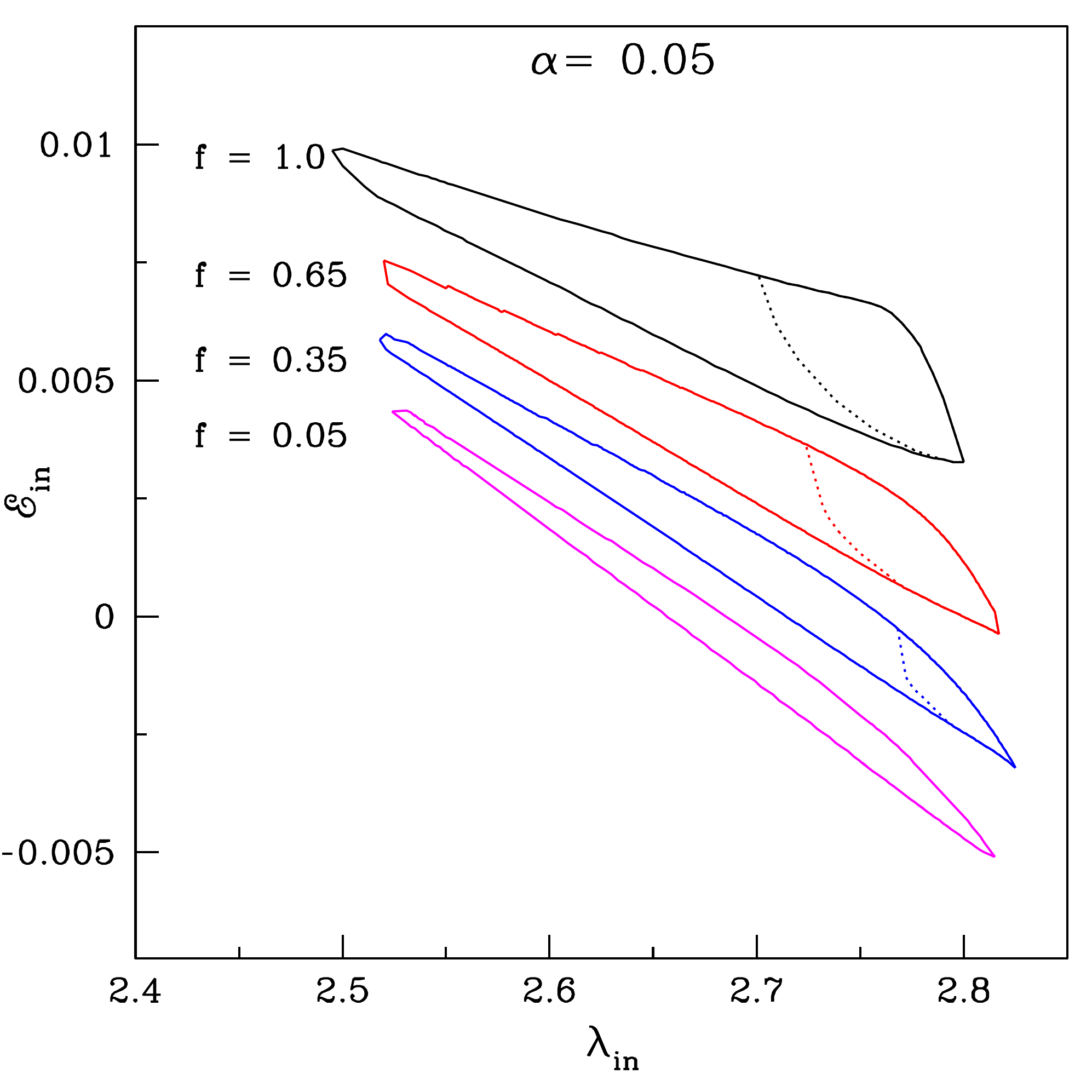}
\end{center}
\caption{
Comparision of shock parameter spaces obtained for various cooling parameters.
Region bounded by the solid curve is for without mass loss case and the region
bounded by the dotted curve represents results with mass loss case. Cooling
parameters are marked in the figure. Here, we consider $\alpha = 0.05$ and $a_k = 0.5$.
See text for details.
}
\end{figure}

We continue our study of shock parameter space in the  $\mathcal{E}_{\rm in} - \lambda_{\rm in}$
plane in presence of cooling. Towards this, we examine the modification of the parameter
space with the increase of the cooling factor $f$ and present it in Fig. 10. 
Here, we choose
$a_k=0.5$ and $\alpha = 0.05$, respectively. As before, the region bounded by the solid
and dotted curves represent the shock  parameter space in absence and in presence of mass
loss. In this figure, chosen cooling factors ($f$) are marked. As the cooling is increased, 
the effective region of the shock parameter spaces both with and without mass loss  is reduced and 
it is shifted towards the negative energy region. Needless to mention that beyond a critical cooling
limit, namely $f\rightarrow 0.05$, parameter space for outflow disappears completely. 

\subsection{Application to source GRO J1655-40}

\begin{figure}
\begin{center}
\includegraphics[angle = 270, width=0.45\textwidth]{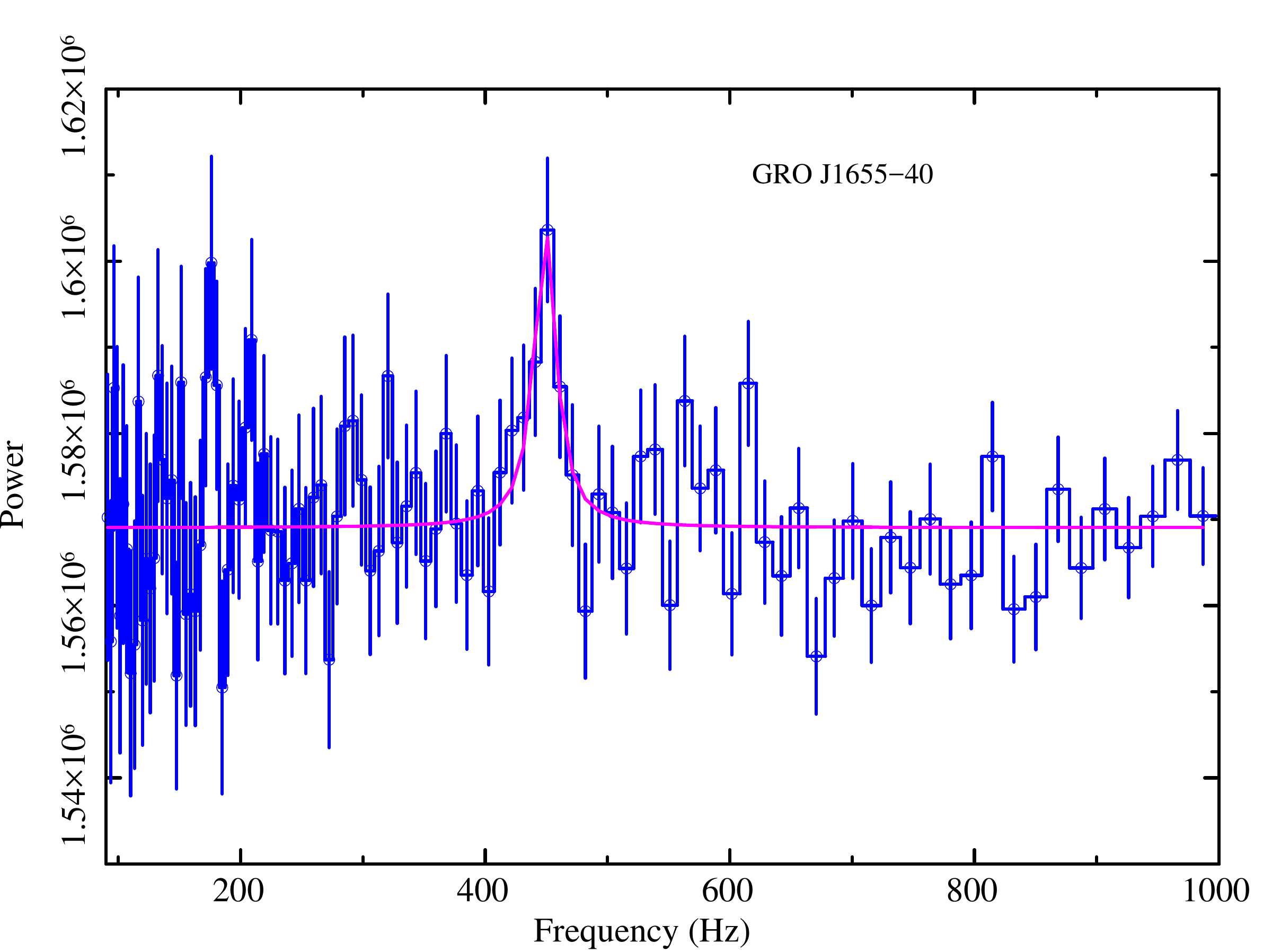}
\end{center}
\caption{Power Density Spectrum (PDS) of the source GRO J1655-40 observed on MJD 50335.9,
showing the HFQPO signature of frequency ($\sim 450~{\rm Hz}$). See text for details.
}
\end{figure}

\begin{figure}
\begin{center}
\includegraphics[angle = 270, width=0.45\textwidth]{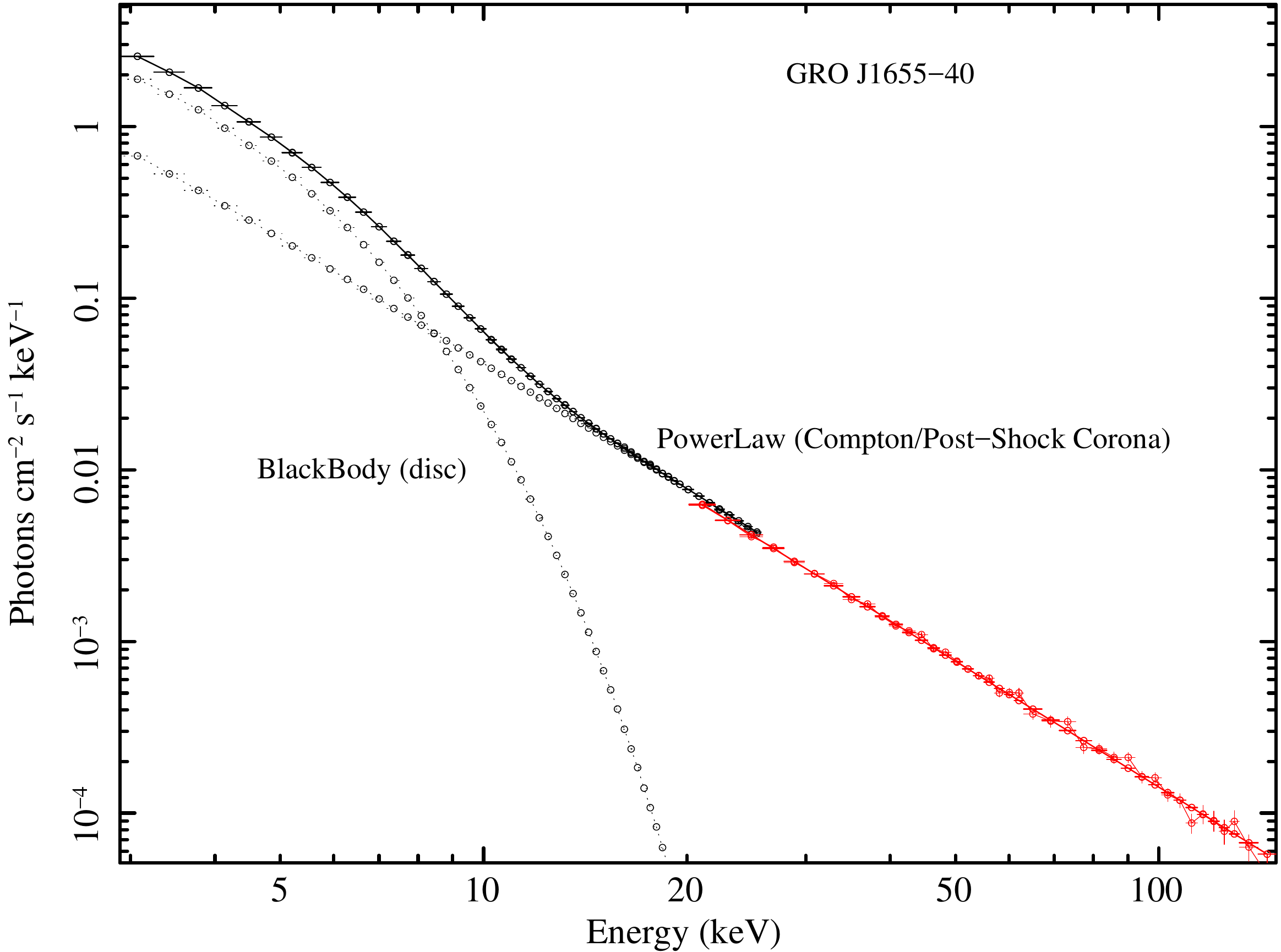}
\end{center}
\caption{X-ray energy spectrum of the black hole source GRO J1655-40 observed on 
MJD 50335.9. Spectrum is fitted with the phenomenological models consisting
of $diskbb$ and $powerlaw$ components. See text for details.
}
\end{figure}

In this section, we attempt to constrain the rotation parameter i.e., spin of 
astrophysical black holes based on our accretion-ejection formalism. For the 
purpose of representation, we consider a Galactic black hole source GRO J1655-40
which is transient in nature and known to produce superluminal jets
\citep{Hjellming-Rupen95, Tingay-etal95, Zhang-etal97}. The source is considered to 
be one of the nearest Galactic black hole source with a distance of 3.2 kpc 
\citep{Migliari-etal07} and inclination angle 69 $\pm$ 2 \citep{Hernandez-etal08}.
So far, it has been observed that this source exhibits the maximum QPO frequency
($\sim$450 Hz) among all the known black hole candidates 
\citep{Remillard-etal02,Belloni-etal12}. The mass of the source is believed to
be well constrained by dynamical method that estimates the range of mass to be
$5.1M_{\odot}$ to $6.3M_{\odot}$ \citep{Greene-etal01,Beer-Podsiadlowski-02}.
However, the measurement of the spin of the black hole remains inconclusive yet.
Numerous  groups used various methods while predicting the spin of the 
black hole with large uncertainties, namely the {\it spectral continuum model} predicts
the spin to be in the range of $0.65$ to $0.75$ \citep{Shafee-etal06}, whereas
\citet{Miller-etal09} predicted the spin to be in the range of $0.94$ to $0.98$ 
using {\it relativistic reflection} and {\it disk continuum emission} models.
However, recent measurement which is based on 
the relativistic precession model of low frequency QPOs obtains the spin to be 
$\sim 0.29$ \citep{Motta-etal14}. Meanwhile, in the theoretical front, efforts were
made to model the origin of the high frequency QPOs considering the coupling
between the angular momentum of the black hole ($i.e.,$ spin) and orbital frequency
at the innermost  stable circular orbit (ISCO) and obtain the range of black hole
spin as $0.2 < a_k < 0.67$ \citep{Abramowicz-Kluzniak01}. A very recent work
carried out by \citet{Stuchlik-Kolos16} predicted the lower limit of the spin
to be $>0.3$ based on the non-geodesic string loop oscillation model of
twin high frequency QPOs. Since the discrepancy of the spin measurement is
not settled yet, this motivates us to constrain the rotation parameter based
on our formalism. 


In this analysis, we choose the maximum observed 
QPO frequency ($\sim$ 450 Hz) for the source as a parameter
($\nu_{QPO}^{\rm max}$) in order to constrain the spin of GRO J1655-40.
Subsequently,  we consider one of the observations of the 1996 
outburst data of the source in order to extract the HFQPO feature and its X-ray spectral nature. 
We used the archival data of {\it RXTE} 
satellite observed on MJD 50335.9 ($09^{\rm th}$ September of 
1996) with typical exposure of $\sim$ 4 ksec \citep{Remillard-etal02,Belloni-etal12}. 
Standard techniques for data reduction and analysis are followed for 
temporal and spectral analysis of {\it RXTE} data.
We extract the lightcurve using the event 
mode data with a time resolution of 0.38 msec. This lightcurve was used to 
generate the power spectrum with 4096 newbins and with a geometrical
binning factor of -1.02. Since we are interested to search for HFQPOs,
we consider only the frequency range of 100 Hz to 1000 Hz. 
Fig. 11 shows the power spectrum plotted in the Lehay power-Frequency 
space. The model used to fit the power spectrum consisted of a {\it constant }
and a {\it Lorentzian}. The best-fitted model parameters are HFQPO $\sim$ 449 
Hz, FWHM $\sim$ 21 Hz with $\chi^{2}_{red}$ of 0.88 ($\chi^{2}$/dof = 85.5/94).
We extract the energy spectrum of the same observation using the standard 
method/techniques \citep{Radhika-etal16a}
from the PCA (3-25 keV) and the HEXTE (20-150 
keV). The combined spectrum (3 -150 keV) was modelled using a 
phenomenological model of {\it diskbb} and {\it powerlaw} for the thermal emission 
from the disk and Comptonized component (i.e., inner part of the disk, here 
PSC) respectively. While modeling, we kept fixed $nH$ $\sim$ $0.7 \times 
10^{22}$ (using Dickey \& Lockman (DL) method$^2$)
 \footnotetext[2]{https://heasarc.gsfc.nasa.gov/cgi-bin/Tools/w3nh/w3nh.pl}
and included an absorption edge around 4 keV. Fig. 12 shows the 
un-folded energy spectrum with the model components. The best-fitted 
model parameters are $T_{in}$ $\sim$ 1.32 keV, $N_{diskbb}$ $\sim$ 967, 
photon index ($\alpha$) $\sim$ 2.43 and $N_{po}$ $\sim$ 11.42 with 
$\chi^{2}_{red}$ of 0.74 ($\chi^{2}$/dof = 64.7/87). Spectral parameters indicate 
that the source was in thermally dominated spectral state with strong emission
at high energy.  Therefore, the possible `accretion' scenario perhaps could be 
that the disk is very close to the black hole with a compact corona ($i. e.$, PSC) to produce 
strong thermal emission (disk contribution) at low energy and power-law distribution 
(Comptonised contribution) at high energies (see \citet{Remillard-McClintock06}). 
Usually, as the post-shock flow remains hot and dense compared to the pre-shock
flow, the cooling rate in the post-shock disc becomes high. Interestingly,
when the infall time in the post-shock region is comparable to the cooling time
scale, shock starts to oscillate \citep{Molteni-etal96b}. In an another effort, 
\citet{Das-etal14} carried out hydrodynamical numerical simulations
of black hole accretion disk and showed that when viscosity exceeds its critical
value, post-shock disc exhibits oscillatory behaviour. Irrespective to the origin,
if PSC starts to oscillate, the source demonstrates QPOs
\citep{Lee-etal11,Nandi-etal12,Sukova-Jianuk15}. Similarly, when the
compact corona exhibits rapid modulation, the source may display high frequency QPOs.

\begin{figure}
\begin{center}
\includegraphics[width=0.45\textwidth]{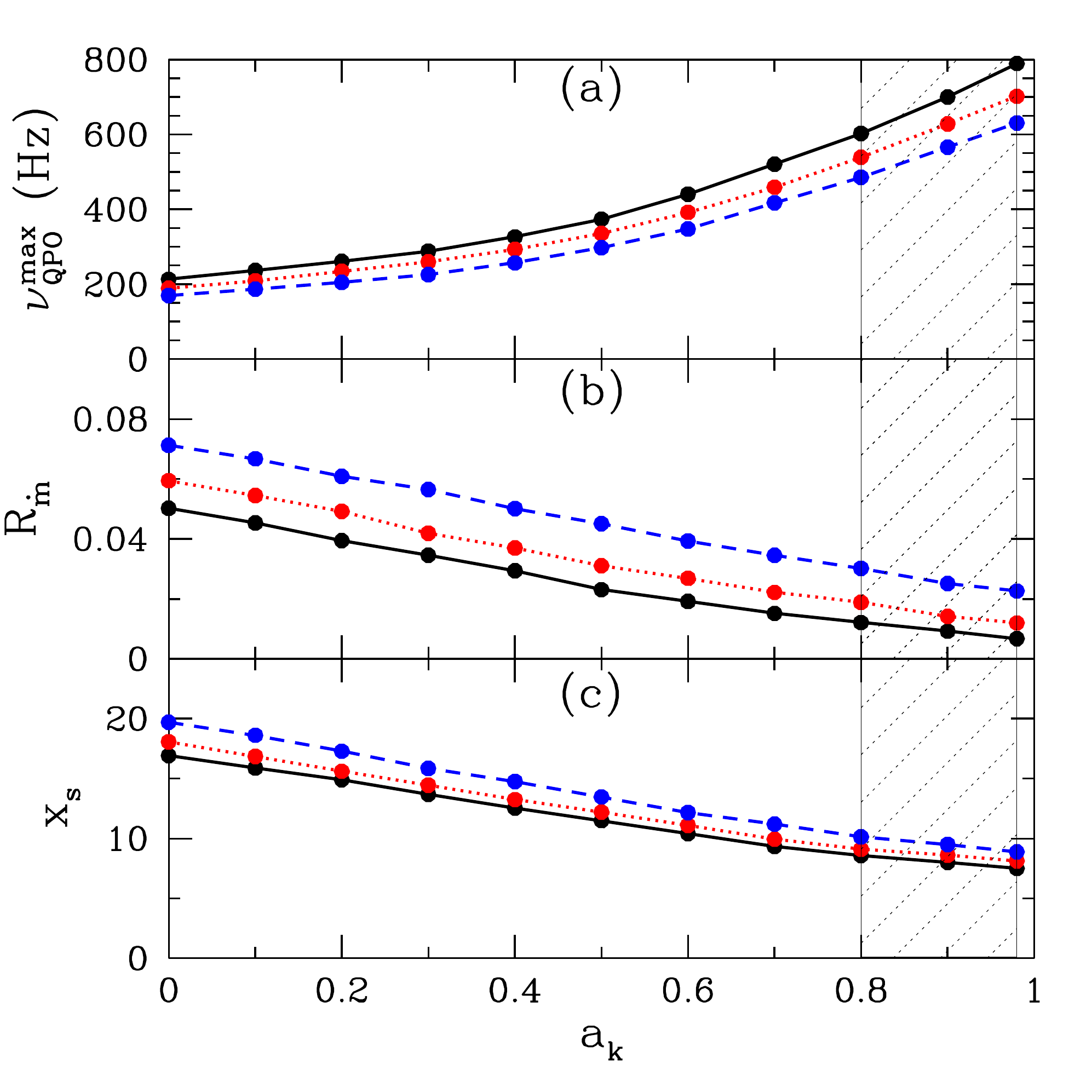}
\end{center}
\caption{
Variation of (a) maximum QPO frequency $(\nu_{QPO}^{\rm max})$, 
(b) corresponding mass outflow rates ($R_{\dot{m}}$) and (c)
shock location ($x_s$) with the spin $(a_k)$ corresponding to
the black hole mass ($M_{\rm BH}=6M_\odot$) of GRO J1655-40. Solid,
dotted and dashed curves represent $\alpha = 0.01$ (black),
$0.05$ (red) and $0.1$ (blue), respectively. In each panel,
shaded region denotes results for $a_k > 0.8$. See text for details.
}
\end{figure}

Now, we calculate the QPO frequency $(\nu_{QPO})$ based 
on our accretion-ejection model. Here, we compute the infall time 
from the post-shock velocity profile as $t_{infall}$ =  $\int{dt}$ = 
$\int_{x_s}^{x_{in}}\frac{dx}{v(x)}$, where $v(x)$ is the post-shock 
velocity. The integration is carried out from the shock location to the inner 
sonic point as the distance between the inner sonic point and event horizon 
is negligibly small. We estimate the QPO frequency as $\nu_{QPO}$ = 
$\frac{1}{t_{QPO}} \sim \frac{1}{t_{infall}}$ in units of $\frac{r_g}{c}$ 
\citep{Molteni-etal96b}. Accordingly, QPO frequency is converted into the unit of Hertz 
when it is multiplied with $\frac{c}{r_g}$. Here, we consider
the mass of the  source GRO J1655-40 as 6$M_{\sun}$. 
With this, we calculate the maximum QPO frequency ($\nu^{\rm max}_{\rm QPO}$)
as function of $a_k$ for various viscosity parameters as depicted in Fig. 13a. Here,
we vary the remaining flow variables ($i. e., {\cal E}_{\rm in}$ and $\lambda_{\rm in}$) 
freely to obtain $\nu^{\rm max}_{\rm QPO}$.
Filled circles connected with the solid, dotted and dashed curves denote
the results corresponding to $\alpha = 0.01$ (black), $0.05$ (red) and $0.1$ 
(blue), respectively. 
In the present analysis, when $a_k$ is increased, shock usually forms
closer to the black hole (see Fig. 3) which apparently 
yields the high frequency QPOs. Therefore, a positive
correlation between $\nu^{\rm max}_{\rm QPO}$
and $a_k$ is very much desirable and it is seen in Fig. 13a. 
Moreover, we find that the maximum QPO frequency generally lies in the range 
$150~{\rm Hz} \lesssim \nu_{QPO}^{\rm max} \lesssim 780~{\rm Hz}$
depending on the values of $a_k$ and $\alpha$ for this source.
The variation of the outflow rates $(R_{\dot{m}})$ corresponding to the
results displayed in the upper panel of Fig. 13 is shown in the middle
panel (Fig. 13b). As  $\nu_{QPO}^{\rm max}$ corresponds to the minimum shock
location ($x_{s}^{\rm min}$) and $\nu_{QPO}^{\rm max}$ increases with the increase
of $a_k$, the size of PSC (which is basically the area of the jet base)
is consequently reduced resulting the lowering of outflow rate.
For completeness, we also show the variation of shock location associated
to $\nu_{QPO}^{\rm max}$ in the bottom panel (Fig. 13c). In each panel, the
shaded region represents the results obtained for $a_k\geq 0.8$.

Since the observed maximum QPO frequency ($\nu_{QPO}^{\rm max}$) of the
source GRO J1655-40 is found as $\sim$ 450 Hz, in our subsequent
analysis, we investigate the ranges of black hole spin parameter ($a_k$) and
viscosity parameter ($\alpha$) that provide $\nu_{QPO}^{\rm max} \sim 450~{\rm Hz}$.
In Fig. 14, we plot $a_k$ along the x-axis and $\alpha$ along the y-axis and
identify the region that renders the $\nu_{QPO}^{\rm max}$ for this source. Based on our model
calculation, we observe that the spin parameter of the source under consideration
has the value $a_k\ge0.57$ which seems to be in agreement with the
observational results of \citet{Shafee-etal06} and \citet{Miller-etal09}.
In addition, a partial agreement is also seen between our estimated range
of spin value and the result reported by \cite{Abramowicz-Kluzniak01}.
In addition, we find that the allowed range of $\alpha$ for this source
is restricted below $\alpha^{\rm max}\sim 0.18$. The correlation study between 
the spin parameters of other black hole sources and their observed
HFQPOs is in progress and will be reported elsewhere.

\begin{figure}
\begin{center}
\includegraphics[width=0.45\textwidth]{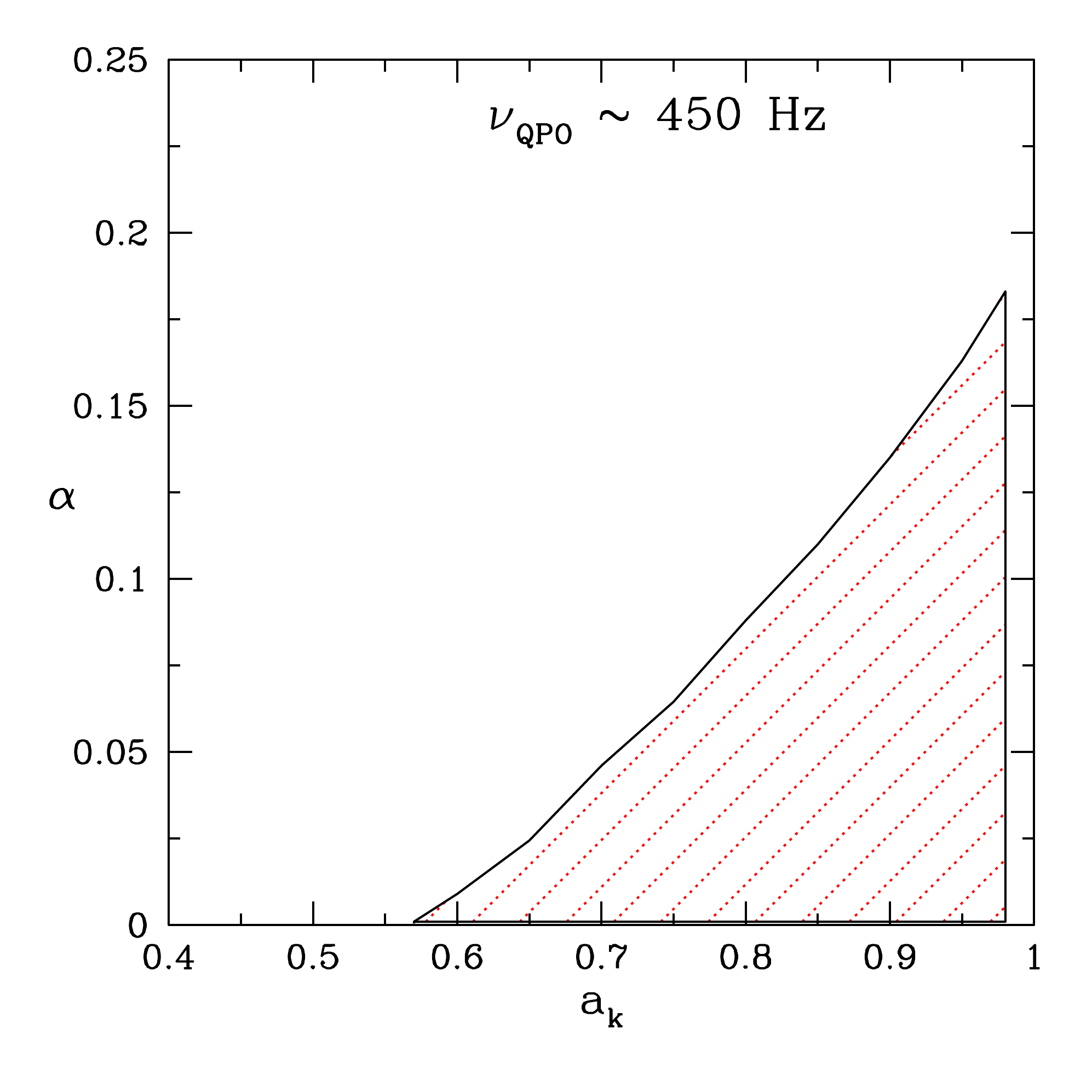}
\end{center}
\caption{Identification of the ranges of viscosity parameter ($\alpha$) and
black hole spin ($a_k$) that cater HFQPO frequency $\nu_{QPO} \sim 450~{\rm Hz}$
observed in GRO J1655-40. See text for details.
}
\end{figure}

\section{Concluding remarks}

In this paper, we describe a self-consistent formalism to calculate the mass
outflow rates from the dissipative accretion flow around the rotating black
holes in presence of viscosity and cooling processes. Effort of this kind was not
considered before to investigate the outflow properties around rotating black
holes. We present the governing
equations for accretion and ejection mechanisms in details and demonstrate
the methodology to obtain the inflow-outflow solutions. Since the black hole
accretion solutions are transonic in nature, we present the sonic point
analysis and demonstrate the sonic point properties. With this, we obtain the 
complete global accretion solutions that may contain shock wave. Because of shock
compression, an excess thermal pressure is developed across the shock that
causes the post-shock flow to become puffed up which equivalently behaves like
Comptonized corona ($i.e.$, PSC). A part of the inflowing matter after being intercepted
at PSC is diverted along the black hole rotation axis to form thermally driven
bipolar outflow (Molteni et al. 1996). Considering this appealing mechanism,
subsequently, we
obtain the coupled inflow-outflow solutions. We show that for a wide range
of inflow parameters, accretion flow can exhibit outflows even when the
viscosity and cooling are very high. We observe that shocks in accretion
flow form within the range of few to several tens of Schwarzschild radius depending
on the inflow parameters. We also observe that as the effect of dissipation
is increased, namely in the form of viscosity and cooling processes, shock moves 
closer to the horizon and the corresponding outflow rate ($R_{\dot m}$)
correlates with the shock location for flows with same outer boundary
condition (Fig. 3).

We show that outflows from the dissipative accretion disc can occur around 
non-rotating as well as rotating black holes. When a part of the inflowing
matter is ejected out from the PSC as outflow, the post-shock pressure is
decreased that eventually allows the shock front to proceed towards the horizon
to maintain the pressure balance across the shock. Interestingly, in absence of
mass loss ($R_{\dot m} = 0$), when the triggering of shock transition takes place
at its minimum location from the horizon ($x^{\rm min}_s$), 
the coupled inflow-outflow solution ceases to exist as the stationary
shock conditions are not satisfied there. 
This eventually indicates that when the effects of viscosity and cooling remain fixed,
the range of the inflow parameters for outflow solutions is reduced compared to the
global shocked accretion solution with $R_{\dot m}=0$. Effectively, 
the domain of the shock parameter space spanned by ${\lambda_{\rm in}}$ and 
${\cal E}_{\rm in}$ is shrunk with the increase of dissipation and it is reduced
further when mass-loss is included (Fig. 5). 
In this study, we observe three distinct features of the modified shock
parameter space. We find that when viscosity parameter ($\alpha$) is increased
keeping spin of the black hole ($a_k$) and cooling parameter ($f$) fixed,
the effective area of the shock parameter space, both in absence and presence
of mass loss, shifts to higher energy and lower angular momentum domain.
This happens due to the fact that the accreting matter becomes relatively
much hotter when it enters into the gravitational potential of rapidly spinning
black hole compared to the non-rotating black hole \citep{Kumar-Chattopadhyay14}.
We further compare the shock parameter spaces for flows with
increasing viscosity parameter keeping cooling parameter ($f$) and $a_k$ fixed.
Due to increase of viscosity, accreting matter not only transports more and more
angular momentum outward, but it also causes the flow to become hotter.
This evidently renders the shock parameter space to shift to the higher
energy and lower angular momentum domain (Fig. 5).
For a cooling dominated flow, the shock parameter space is shrunk and
shifted to negative energy region (Fig. 10).

We estimate the maximum viscosity parameter that provides global shocked
accretion solution both by excluding mass loss ($\alpha^{\rm max}_{no}$) and
including mass loss ($\alpha^{\rm max}_{o}$). Since the possibility of shock formation
reduces in presence of mass loss, we obtain $\alpha^{\rm max}_{o} < \alpha^{\rm max}_{no}$
for a given $a_k$. We further find that the obtained values of $\alpha^{\rm max}_{no} \sim 0.35$
and $\alpha^{\rm max}_{o} \sim 0.25$ in the limit of $a_k \rightarrow 0$ are in agreement
with the results reported by \citet{Chakrabarti-Das04} and \citet{Kumar-Chattopadhyay13}.
Moreover, we calculate the maximum mass outflow rate ($R^{\rm max}_{\dot m}$) as function
of viscosity parameter ($\alpha$) by exploring all possible combination of inflow parameters
and find that $R^{\rm max}_{\dot m}$ is largely obtained for higher ${\cal E}_{\rm in}$ and lower 
$\lambda_{\rm in}$ values irrespective to the values of $a_k$ (Fig. 7).
In Fig. 8, we depict the
computed values of $R^{\rm max}_{\dot m}$ as function of $\alpha$ and find the
range as $3\% \le R^{\rm max}_{\dot m} \le 19\%$. 

It is worthy to address a very interesting finding that is emerged out from 
our study here. We have pointed out that when the viscosity as well as cooling
parameters are considered beyond their critical values, coupled inflow-outflow
solution ceases to exist. However, non-steady shock still may present as
shown by \citet{Ryu-etal97,Das-etal14}.
Meanwhile, \citet{Molteni-etal96b,Chakrabarti-Manickam00} showed
that QPO frequency of emergent hard radiation from the black holes is proportional
to the infall time of the accreting matter. Adopting this prescription, we employ 
our formalism to calculate the maximum QPO frequency ($\nu^{\rm max}_{\rm QPO}$)
as function of $a_k$ for flows with various $\alpha$. We then attempt to constrain
the spin of the black hole source GRO J1655-40 considering the highest QPO 
frequency $\sim 450$ Hz observed in this source (Fig. 11). 
Based on our present analysis, we find that $a_k \ge 0.57$ for this source
(Fig. 14).

An important point is to be noted that 
in late 90's, \citet{Blandford-Begelman99} found the
advection dominated inflow-outflow solutions (ADIOS) around a Newtonian central 
object. There are subtle differences between our study and ADIOS, one of which arises
mainly due to the choice of the accretion rate equation. In ADIOS model, the accretion
rate is assumed to satisfy the power low dependence on radial coordinate as 
${\dot M} \propto x^s$ with $0 \le s < 1$ whereas no such restriction is imposed in
our study (see equation (3)). In addition, in this work, we have considered
dissipative accretion flow including viscosity and radiative cooling which were
ignored in ADIOS model. Moreover, we obtain the self-consistent 
accretion-ejection solutions that are coupled via Rankine-Hugoniot shock
and in ADIOS model, mass loss is allowed to take place from all radial coordinates.

In this work, the relativistic effects of the central black hole are taken in to account
by adopting the pseudo-Kerr gravitational potential \citep{Chakrabarti-Mondal06}.
This potential satisfactorily describes the general relativistic effect around black hole
for $a_k\le0.8$ and it allows us to investigate properties of coupled inflow-outflow
solutions in a simpler way. However, we continue our study even for  $a_k > 0.8$ to
oversee the overall properties of the inflow-outflow solutions. In addition, for simplicity, we
consider the parametric cooling mechanism ignoring the physically motivated
cooling processes. Moreover, we consider the constant adiabatic index to describe
the accretion-ejection model instead of estimating it self-consistently based on the
thermal properties of the flow. In this formalism, we have disregard 
the issues of the collimation of jets and its powering processes although magnetic 
fields may be an important ingredient to explain the transient relativistic jets and its
collimation mechanism. Implementation of all such relevant issues are beyond
the scope of the present paper. However, we argue that even with the above
approximations, overall findings of our present analysis will remain qualitatively unaltered.

\section*{Acknowledgments}
Authors thank the anonymous referee for useful comments
and constructive suggestions.
AN thanks GD, SAG; DD, PDMSA and Director, ISAC for encouragement and
continuous support to carry out this research. We thank Toru Okuda
for discussion. This research has made use of the data obtained
through High Energy Astrophysics Science Archive Research Center
on-line service, provided by NASA/Goddard Space Flight Center.

\label{lastpage}

\end{document}